\documentclass[twoside,twocolumn,9pt,abstract=on]{article}

\usepackage{booktabs}
\usepackage{extsizes}
\usepackage[super,sort&compress,comma]{natbib} 
\usepackage[version=3]{mhchem}
\usepackage[left=1.5cm, right=1.5cm, top=1.785cm, bottom=2.0cm]{geometry}
\usepackage{balance}
\usepackage{times}
\usepackage{times,newtxmath} 
\usepackage{sectsty}
\usepackage{graphicx} 
\usepackage{lastpage}
\usepackage[format=plain,justification=raggedright,singlelinecheck=false,font={stretch=1.125,small,sf},labelfont=bf,labelsep=space]{caption}
\usepackage{float}
\usepackage{fancyhdr}
\usepackage{fnpos}
\usepackage[english]{babel}
\usepackage{array}
\usepackage{droidsans}
\usepackage{charter}
\usepackage[T1]{fontenc}
\usepackage[usenames,dvipsnames]{xcolor}
\usepackage{setspace}
\usepackage[compact]{titlesec}
\usepackage{bm}
\usepackage{bbm}
\usepackage{environ}
\usepackage{hyperref}	

\usepackage{authblk}

\usepackage{amssymb}
\usepackage{color}
 \usepackage{soul} % para tachar palabras
\usepackage{afterpage}
\usepackage{multicol}

\usepackage{epstopdf}%This line makes .eps figures into .png - please comment out if not required.

\title{Spin transport in graphene/transition metal dichalcogenide heterostructures}
\author[1,4]{Jose H Garcia}
\author[1,2]{Marc Vila}
\author[1]{Aron W Cummings}
\author[1,3]{Stephan Roche}
\affil[1]{Catalan Institute of Nanoscience and Nanotechnology (ICN2), CSIC and BIST, Campus UAB, Bellaterra, 08193 Barcelona, Spain.}
\affil[2]{Department of Physics, Universitat Aut\`onoma de Barcelona, Campus UAB, Bellaterra, 08193 Barcelona, Spain.}
\affil[3]{~ ICREA - Instituci\'o Catalana de Recerca i Estudis Avan\c{c}ats, 08010 Barcelona, Spain.}
\affil[4]{~Corresponding author:  \href{mailto:josehugo.garcia@icn2.cat}{josehugo.garcia@icn2.net}}
\date{}                     %% if you don't need date to appear
\begin{document}

\twocolumn[
  \begin{@twocolumnfalse}
    \maketitle
    \begin{abstract}
Since its discovery, graphene has been a promising material for spintronics: its low spin-orbit coupling, negligible hyperfine interaction, and high electron mobility are obvious advantages for transporting spin information over long distances. However, such outstanding transport properties also limit the capability to engineer active spintronics, where strong spin-orbit coupling is crucial for creating and manipulating spin currents. To this end, transition metal dichalcogenides, which have larger spin-orbit coupling and good interface matching, appear to be highly complementary materials for enhancing the spin-dependent features of graphene while maintaining its superior charge transport properties. In this review, we present the theoretical framework and the experiments performed to detect and characterize the spin-orbit coupling and spin currents in graphene/transition metal dichalcogenide heterostructures. Specifically, we will concentrate on recent measurements of Hanle precession, weak antilocalization and the spin Hall effect, and provide a comprehensive theoretical description of the interconnection between these phenomena. 

{\bf This document is the unedited Author's version of a Submitted Work that was subsequently accepted for publication in Chemical Society Reviews, copyright Royal Society of Chemical after peer review. To access the final edited and published work see} \href{http://pubs.rsc.org/en/Content/ArticleLanding/2018/CS/C7CS00864C}{ this link}1

    \end{abstract}
  \end{@twocolumnfalse}
]
\maketitle

%Please use \dag to cite the ESI in the main text of the article.
%If you article does not have ESI please remove the the \dag symbol from the title and the footnotetext below.
%-\footnotetext{\dag~Electronic Supplementary Information (ESI) available: [details of any supplementary information available %should be included here]. See DOI: 10.1039/b000000x/}
%additional addresses can be cited as above using the lower-case letters, c, d, e... If all authors are from the same address, no letter is %required

%%%END OF FOOTNOTES%%%

%%%MAIN TEXT%%%%

%%%%%%%%%%%%%%%%%%%SECTION%%%%%%%%%%%%%%%%%
\section{Introduction}
%%%%%%%%%%%%%%%%%%%%%%%%%%%%%%%%%%%%%%%%
Graphene, a two dimensional carbon allotrope, has proven over the last decade to be a truly wonder material. Despite its simple structure and chemical composition, its use in a variety of fields such as coatings, sensors, energy, biomedicine, photonics, and optoelectronics has demonstrated the richness of its properties and its potential industrial impact \cite{Peng2014, Ferrari2015, Yin2016, Sun2015, Mao2017, Reina2017}. The usefulness of graphene for spintronics has also been recognized, owing to its high electron mobility \cite{Bolotin2008, Banszerus2015, Banszerus2016}, small spin-orbit coupling \cite{Min2006, Huertas2006, Gmitra2009, Konschuh2010} and negligible hyperfine interaction \cite{Wojtaszek2014, Han2014}. All together these factors result in coherent spin propagation over unprecedented microscopic distances \cite{Guimaraes2014, Drogeler2014, Kamalakar2015, Singh2016, Drogeler2016, Drogeler2017}, proving graphene to be a suitable enabling material for passive spin components \cite{Roche2014, Roche2015}. However, the features that make graphene a good transporter of spin also make it a poor candidate for active spintronics devices, which require the generation or manipulation of spin currents. Spin-orbit coupling (SOC), a relativistic phenomenon that couples an electron's spin with its momentum, is one method for realizing active spintronics. Although SOC is very small in carbon-based materials, there is an increasing body of evidence showing that it can be artificially enhanced in graphene, opening a path for the implementation of active graphene-based spintronic devices \cite{CastroNeto2009, Weeks2011, Gmitra2013, Cresti2014, Ferreira2014, Balakrishnan2014, Gmitra2016, Alsharari2016, Cummings2017, Avsar2014, Wang2015, Wang2016, Yang2016, Yan2016, Yang2017, Ghiasi2017, Benitez2017, Omar2017, Omar2017b, Dankert2017, Volkl2017, Wakamura2017, Zihlmann2018, Friedman2018}.

The SOC in graphene can be enhanced by chemical functionalization, or by placing it on a high-SOC substrate. Between these two approaches, high-SOC substrates are preferred because they are chemically inert and electrically insulating, and thus will have little impact on graphene's electronic properties. Meanwhile, weak hybridization with the substrate can enhance the SOC in the graphene layer, potentially leading to predicted phenomena such as the spin Hall effect (SHE) \cite{Ferreira2014, Pachoud2014, Huang2016, TienVo2017, Dyrda2017, Milletari2017} or the quantum spin Hall effect (QSHE) \cite{KaneMele2005, Weeks2011}. Experiments carried out over the past few years have reported strong evidence of such proximity-induced SOC enhancement in graphene when interfaced with transition metal dichalcogenides (TMDCs), leading to a three order of magnitude increase of SOC and the proposal of new spin device functionalities \cite{Avsar2014, Wang2015, Wang2016, Yang2016, Yan2016, Yang2017, Ghiasi2017, Benitez2017, Omar2017, Omar2017b, Dankert2017, Volkl2017, Wakamura2017, Zihlmann2018, Friedman2018}. Therefore, an in-depth understanding and characterization of the properties of graphene/TMDC heterostructures appears to be fundamental for the field of spintronics.

The materials used in spintronics applications can be characterized by three fundamental figures of merit: the spin relaxation time $\tau_\text{s}$, the spin diffusion length $\lambda_\text{s}$, and the spin Hall angle $\gamma_\text{sH}$. The relaxation time dictates the upper time limit within which spin information can be transmitted and manipulated, hence large $\tau_\text{s}$ is desired. The spin diffusion length denotes the distance over which spin currents can propagate without losing information, and must also be maximized. Both quantities are related by the spin diffusion coefficient $D_\text{s}$, which depends on the sample mobility and the transport regime, and is given in the diffusive regime by $\lambda_\text{s} = \sqrt{D_\text{s} \tau_\text{s}}$. Finally, the spin Hall angle, which is the figure of merit of the spin Hall effect \cite{RevModPhys.87.1213}, measures the efficiency of charge-to-spin conversion and vice versa. This quantity must be maximized for practical use in data storage or future non-charge-based information processing technologies. These parameters are usually evaluated through three different types of experiments: (i) measurements of Hanle precession in lateral spin valves, (ii) measurements of the weak antilocalization effect, and (iii) measurements of magnetic-field-modulated nonlocal resistance in Hall bar geometries. Each of these approaches yields some combination of $\tau_\text{s}$, $\lambda_\text{s}$, and $\gamma_{\text{sH}}$, but the obtained values can differ depending on the measurement technique. This has led to some apparent contradictions when compared to the available theoretical framework, and calls for a better understanding of these phenomena and the relationship between them in order to move forward into practical applications.

In this review, we present a unified description of graphene/TMDC heterostructures in the context of spintronics, through a discussion of some of the most relevant experiments and their relationship to recent theoretical developments. Unless stated otherwise, we will focus exclusively on monolayer graphene since it is the material explored in the majority of experiments and theoretical works. The review is organized as follows: Section \ref{pristine_graphene} starts with an introduction to spin transport measurements performed in graphene supported on silicon oxide and hexagonal boron nitride (hBN) substrates. Additionally, we give an overview of the most relevant evidence of spin-orbit coupling enhancement in graphene and how it was determined. We also include a brief review of graphene/TMDC devices. These discussions will serve as a baseline for understanding new results in graphene/TMDC systems. In Section \ref{sec_theory} we introduce the theory of spin-orbit coupling and spin relaxation in graphene/TMDC heterostructures, which will be used to interpret and analyze the measurements discussed in the following sections. In Section \ref{spin-precession}, we discuss the theory of Hanle precession in lateral spin valves and its extension to anisotropic systems. We then use the theoretical framework presented in Section \ref{sec_theory} and the Hanle precession theory to offer some insights into recent experimental results. In Section \ref{sec_wal} we review weak antilocalization (WAL) theory, summarize the measurements of WAL in graphene/TMDC heterostructures, and comment on the need to consider valley-Zeeman SOC in the analysis of these measurements. We also discuss the limits of traditional WAL analysis in systems with very strong SOC. In Section \ref{sec_she} we present a basic introduction of the spin Hall effect and we discuss the complexity of the phenomenon in graphene/TMDC heterostructures. Then we discuss numerical predictions of the SHE in these systems and propose some methods for its experimental observation. The Rashba-Edelstein effect is also briefly discussed. The review concludes in Section \ref{sec_conclusions}, where we highlight the main take-home messages and offer some perspectives for the future of these devices, as well as for graphene interfaced with other materials.

%%%%%%%%%%%%%%%%%%%SECTION%%%%%%%%%%%%%%%%%
\section{History of spin transport in graphene} \label{pristine_graphene}
%%%%%%%%%%%%%%%%%%%%%%%%%%%%%%%%%%%%%%%%

%%%%%%%%%%%%%%%%%SUBSECTION%%%%%%%%%%%%%%%%%
\subsection{Graphene on traditional substrates}
%%%%%%%%%%%%%%%%%%%%%%%%%%%%%%%%%%%%%%%%
Owing to its small spin-orbit coupling ($\sim$$\upmu$eV) and hyperfine interaction \cite{Min2006, Huertas2006, Gmitra2009, Konschuh2010, Wojtaszek2014, Han2014}, graphene is expected to possess long spin relaxation times $\tau_\text{s}$ and lengths $\lambda_\text{s}$. Indeed, the first theoretical studies predicted $\tau_\text{s}$ exceeding the $\upmu$s range \cite{Ertler2009, Huertas2009, Zhou2010}, which is orders of magnitude larger than in typical metals and semiconductors \cite{Zutic2004}. These estimates were based on traditional mechanisms of spin relaxation in metals and semiconductors, namely the Elliott-Yafet (EY) and D'yakonov-Perel' (DP) mechanisms. In the EY mechanism \cite{Elliott1954, Yafet1963}, an electron's spin has a finite probability of flipping during a scattering event, leading to a spin relaxation time that is proportional to the momentum scattering time $\tau_\text{p}$. In graphene this relation is given as $\tau_\text{s} \sim (E_\text{F} / \lambda)^2 \tau_\text{p}$, where $\lambda$ is the SOC strength and $E_\text{F}$ is the Fermi energy \cite{Ochoa2012}. In the DP mechanism \cite{dyakonov1971spin}, SOC leads to electron spin precession and dephasing between scattering events. When the scattering time is shorter than the precession time, electrons tend to maintain their spin orientation, in what is known as motional narrowing. This leads to a spin relaxation time that is inversely proportional to the momentum scattering time, $\tau_\text{s} \sim (\hbar / \lambda)^2 / \tau_\text{p}$, where $\hbar$ is the reduced Planck constant. The typical scattering events considered in the calculations are collisions with charged impurities \cite{Ertler2009, Zhou2010}, ripples \cite{Huertas2006, Huertas2009, Fratini2013, Vicent2017}, phonons \cite{Ertler2009, Zhou2010, Ochoa2012phonon, Fratini2013, Vicent2017} and other electrons \cite{Zhou2010}. In combination with the small SOC in graphene, these mechanisms are predicted to give very long spin relaxation times \cite{Han2014}.

The nonlocal Hanle spin precession measurement is the typical experimental technique employed to study spin relaxation in graphene, as it allows the extraction of both $\tau_\text{s}$ and $\lambda_\text{s}$ \cite{Jedema2002, ActaFabian, Wu2010}. Early measurements of graphene spin valves revealed $\tau_\text{s} < 1$ ns \cite{Tombros2007, Tombrosa2008, Jozsa2009, Han2011, Avsar2011, Zomer2012, Guimaraes2012, Neumann2013}, while the use of hBN as a substrate or protective layer, in addition to other improvements of device quality, have yielded spin lifetimes up to 12 ns \cite{Guimaraes2014, Drogeler2014, Kamalakar2015, Singh2016, Drogeler2016, Drogeler2017}. Although these results already make clean graphene a suitable platform for achieving longer coherent spin propagation than in typical metals or semiconductors, $\tau_\text{s}$ remains several orders of magnitude below the initial theoretical predictions \cite{Ertler2009, Huertas2009, Zhou2010}, a puzzling result which remains open to discussion and interpretation \cite{Roche2014, Roche2015}. Indeed, this intriguing difference between theory and experiment initially suggested that the DP and EY mechanisms may not be fully appropriate for graphene.

As a result, mechanisms different than spin flip (EY) or spin dephasing (DP) due to SOC and scattering have been proposed to explain this discrepancy. The role of the hyperfine interaction between the nucleus and electrons has been proven negligible because of the low abundance of $^{13}$C and the weak hyperfine coupling in graphene \cite{Wojtaszek2014}. Meanwhile, a well-studied extrinsic source of relaxation comes from the ferromagnetic contacts of the devices \cite{Popinciuc2009, Volmer2013, Drogeler2014}. Experiments have shown that a high contact resistance (i.e., a tunnel barrier) between the contact and graphene enhances the spin lifetime, but only up to a few ns. Theoretical models to account for contact-induced spin relaxation have been developed \cite{Maassen2012, Sosenko2014, Idzuchi2015, Amamou2016, Stecklein2016}, but spin lifetimes of the order of ns are still obtained when contact effects are accounted for. Ripples or corrugation in graphene were also proposed as a source of spin relaxation, where local curvature induces an effective spin-orbit field, or gauge field \cite{Huertas2009, Ochoa2012phonon, Vicent2017}. However, calculations of this phenomenon in experimentally-relevant situations revealed spin lifetimes in the range of hundreds of ns up to $\upmu$s, indicating that this mechanism is not likely to be a limiting factor in experiments to date.

Another source of spin relaxation, scattering by magnetic impurities, has been experimentally \cite{Lundeberg2013} and theoretically \cite{Kochan2014, Soriano2015} studied. During the device fabrication process, chemical impurities such as hydrogen or other hydrocarbons can be deposited on graphene, and these impurities can give rise to localized magnetic moments \cite{Yazyev2007, Santos2012, Herrero2016}. At the resonant energies of the impurity, the electron spin interacts with the local magnetic moment of the scatterer and acquires high probability to spin-flip. This mechanism predicts that the spin relaxation will be fastest near the graphene Dirac point and is proportional to the density of magnetic impurities, with simulations revealing spin lifetimes on the order of those measured experimentally \cite{Kochan2014, Soriano2015}.

Another spin relaxation mechanism, known as spin-pseudospin coupling, arises in the presence of Rashba SOC, which entangles graphene's spin and pseudospin (or sublattice) degrees of freedom \cite{Tuan2014}. The entanglement of these two degrees of freedom leads to enhanced spin dephasing and relaxation near the graphene Dirac point, qualitatively similar to the case of magnetic impurities. Simulations of this mechanism in the diffusive transport regime also yielded spin lifetimes similar to the experimental values \cite{VanTuan2016}. The effect of magnetic impurities can be avoided by employing different fabrications steps \cite{Drogeler2016}, while spin-pseudospin-induced relaxation is an intrinsic property of graphene that could represent the upper limit of spin lifetime in the ultraclean limit \cite{Cummings2016}.

To determine which mechanism is responsible for spin relaxation, experiments typically probe the dependence of $\tau_\text{s}$ and $\lambda_\text{s}$ on the applied gate voltage, electron density, sample quality, or measurement temperature \cite{Avsar2011, Han2014}. Unfortunately, it remains difficult to unambiguously differentiate one relaxation mechanism from another, given the complex relation between spin dynamics, disorder, and substrate-induced spin-orbit effects \cite{Roche2014}. For this reason, recent studies have focused on the evaluation of spin lifetime anisotropy $\zeta$, defined as the ratio of the lifetime of spins pointing out of the graphene plane to those pointing in the plane. Different spin relaxation mechanisms yield different anisotropy values, making this quantity a useful probe of the nature of the spin dynamics in a given system. For example, the DP mechanism driven by Rashba SOC in disordered two-dimensional systems \cite{ActaFabian} yields $\zeta = 1/2$, while magnetic impurities \cite{Kochan2014, Soriano2015} and the spin-pseudospin coupling mechanism in the clean limit \cite{Tuan2014, VanTuan2016} both produce $\zeta = 1$. Meanwhile, spin relaxation owing to strain-induced gauge fields \cite{Huertas2009} is predicted to yield $\zeta > 1$. Recent measurements of graphene on SiO$_2$ substrates \cite{Raes2016, Raes2017, Guimaraes2014, Tombros2008, Ringer2017} have revealed $\zeta \approx 1$. This rules out the DP mechanism, as well as strain and corrugation, as the main drivers of spin relaxation in these measurements and further supports magnetic impurities or spin-pseudospin coupling. Recent work has shown that by protecting the graphene layer from chemical solvents during device fabrication, the spin lifetime can be increased by more than one order of magnitude \cite{Drogeler2016}, from under 1 ns up to more than 10 ns. This lends strong support to the theory that magnetic impurities are limiting $\tau_\text{s}$ in typical graphene devices, while spin-pseudospin coupling may remain a limiting factor in cleaner devices. Meanwhile, we will show later that anisotropy measurements are key to characterize the proximity-induced SOC in graphene/TMDC heterostructures.

%%%%%%%%%%%%%%%%%%%SUBSECTION%%%%%%%%%%%%%%%%%
\subsection{Enhancement of SOC in graphene}
%%%%%%%%%%%%%%%%%%%%%%%%%%%%%%%%%%%%%%%%

Given the potential technological and fundamental impact, over the years different strategies have been proposed and realized for enhancing the SOC in graphene. The first theoretical proposals relied on the presence of hydrogen or heavy metal impurities, which induce a strong localized enhancement of the SOC in graphene \cite{CastroNeto2009, Weeks2011, Gmitra2013}. Heavy metal impurities were predicted to induce the QSHE in graphene \cite{Weeks2011, Cresti2014}, but experimentally this phenomenon has yet to be observed, owing to clustering of the impurities or other more subtle effects \cite{Cresti2014, Wang2015c, Jia2015, dosSantos2017}. Meanwhile, hydrogen impurities were predicted to induce a giant SHE via the skew scattering mechanism \cite{Ferreira2014}. This prediction was supported by measurements of weakly-hydrogenated graphene in a Hall bar geometry. In these measurements, the appearance of a large nonlocal charge signal was attributed to a hydrogen-induced spin Hall effect, and under this assumption a three-order-of-magnitude enhancement of SOC was reported \cite{Balakrishnan2014}. Although these results agree quite well with initial theoretical predictions \cite{Gmitra2013, Ferreira2014}, to date the experiment has been difficult to reproduce by other groups \cite{Wang2015b, Kaverzin2015}. One problem is that this measurement approach can lead to false positives because a variety of phenomena unrelated to spin may contribute to the nonlocal signals \cite{VanTuan2016-PRL, CrestiRNC2016, Ribeiro2017, Marmolejo-Tejada2017}. Another issue is that resonant scatterers such as hydrogen induce magnetic moments in graphene, which subsequent calculations predict to dominate over any induced SOC \cite{Kochan2014, Soriano2015}.

As mentioned above, it has proven difficult to controllably enhance the SOC in graphene via the adsorption of impurities. In addition to clusterization and the dominance of magnetic moments, impurities also act as scattering centers which degrade graphene's exceptional charge transport properties. For this reason, a great deal of focus has turned to interfacing graphene with high-SOC insulators, where it might be possible to induce a uniform SOC in graphene while maintaining its high electrical mobility. This approach has proven to be especially successful in graphene/TMDC heterostructures, as a large number of works have now demonstrated enhanced SOC in graphene in contact with a variety of TMDCs while maintaining very good charge transport \cite{Avsar2014, Wang2015, Wang2016, Yang2016, Yan2016, Yang2017, Ghiasi2017, Benitez2017, Omar2017, Omar2017b, Dankert2017, Volkl2017, Wakamura2017, Zihlmann2018, Friedman2018}. In the majority of these measurements, the presence of strong proximity-induced SOC was confirmed through the analysis of weak antilocalization, which permits the extraction of the spin relaxation times in the material \cite{Mccann2012}. Spin-orbit strengths can then be evaluated by linking these relaxation times to an assumed theory of spin relaxation, and most analyses have assumed the traditional EY and DP mechanisms. However, recent theoretical developments have indicated the importance of a new method of spin relaxation, driven by a spin-valley coupling through a SOC termed valley-Zeeman \cite{Cummings2017}. As discussed in Sections \ref{sec_theory}-\ref{sec_wal}, this can lead to a giant spin lifetime anisotropy in the graphene layer \cite{Cummings2017}, a result that has been confirmed by recent Hanle measurements of this phenomenon \cite{Ghiasi2017, Benitez2017}, as well as the most recent WAL analysis \cite{Zihlmann2018}. Beyond impacting the features of spin relaxation in graphene, TMDCs are also predicted to significantly enhance phenomena that are directly relevant for low-power applications, namely the spin Hall effect and the Rashba-Edelstein effect \cite{Milletari2017, Offidani2017}. These phenomena are discussued in detail in Section \ref{sec_she}.

%%%%%%%%%%%%%%%%%%%SUBSECTION%%%%%%%%%%%%%%%%%
\subsection{Graphene/TMDC devices} \label{sec_devices}
%%%%%%%%%%%%%%%%%%%%%%%%%%%%%%%%%%%%%%%%

When fabricating graphene/TMDC devices, the quality of the interface becomes an essential aspect of the resulting device properties. A high quality interface will enable a uniform enhancement of the SOC in graphene, while maintaining its high electron mobility. There are different fabrication methods for two-dimensional heterostructures\cite{Frisenda2018}, although for graphene/TMDC devices, the usual fabrication process involves the dry transfer of graphene or the TMDC onto a Si/SiO$_2$ substrate, followed by dry transfer of the other material on top \cite{Avsar2014, Wang2015, Wang2016, Omar2017b, Volkl2017, Friedman2018, Yang2016, Yan2016, Dankert2017, Omar2017, Yang2017, Ghiasi2017, Benitez2017, Wakamura2017, Zihlmann2018}. The graphene and TMDCs layers are usually mechanically exfoliated from their bulk counterparts, although in some cases the they can be grown directly on the Si/SiO$_2$ substrate \cite{Wakamura2017, Friedman2018}. After layer transfer, devices are patterned, then the contacts are deposited and thermal annealing is often used between transfer steps and after device fabrication. In general, this fabrication process can yield good devices, but imperfections such as wrinkles in the graphene, bubbles in the graphene/TMDC interface, and adsorbates on the graphene layer can remain at the end of the fabrication process.

Charge transport measurements are usually used as a global measure of the quality of the graphene layer. Following the above fabrication procedure, electron mobilities in the range of a few thousand up to a few tens of thousands of cm$^2$/V$\cdot$s are now routinely obtained, which are comparable to values reported in graphene on traditional substrates \cite{Guimaraes2014, Drogeler2014, Kamalakar2015, Singh2016, Drogeler2016, Drogeler2017, Tombros2007, Tombrosa2008, Jozsa2009, Han2011, Avsar2011, Zomer2012, Guimaraes2012, Neumann2013}. Thus, charge transport in the graphene layer appears to be largely unaffected by the proximity of the TMDC. In fact, one comprehensive study of these systems found that graphene's electronic properties were greatly improved when encapsulated between different 2D insulators \cite{Kretinin2014}. This was explained by a self-cleaning process where some adatoms and imperfections were accumulated in large pockets at the interface between the layers, thus reducing the scattering and increasing the mobility.

Additionally, with some extra care extremely high mobilities can be obtained. When patterning devices after material transfer, wrinkles or other structural defects in the graphene layer can be minimized. Adsorbates on the graphene layer typically consist of polymer residue accumulated during the transfer process, and can be at least partially removed through thermal annealing. Adsorbates can also be removed by dragging an atomic force microscope (AFM) across the graphene surface; this ``ironing'' technique has resulted in very clean devices with electron mobilities up to $160\,000$ cm$^2$/V$\cdot$s \cite{Wang2016}. Encapsulation with hBN can protect the device from chemical contamination, also yielding mobilities greater than $100\,000$ cm$^2$/V$\cdot$s \cite{Volkl2017, Zihlmann2018}. These results suggest that chemical adsorbates, which are deposited during the fabrication process, are the likely dictate the upper limit of ballistic charge transport in these devices.

In order to induce strong SOC in the graphene layer it is very important to have the highest possible interface quality and close contact with the TMDC, as {\it ab initio} simulations indicate that the strength of the induced SOC decreases exponentially with the interlayer distance \cite{Gmitra2016, Wang2015, Yang2016}. For this reason, bubbles at the graphene/TMDC interface, which are pockets of large separation between the two materials, should be avoided. These bubbles can be detected with photoluminescence (PL) mapping of the sample area \cite{Yang2017}. Owing to their direct band gap, single-layer TMDCs give rise to a strong PL response, but this response is suppressed when the TMDC is in contact with graphene \cite{Shih2014, Defazio2016}. Thus, bubbles should appear as localized regions of PL in the graphene/TMDC device. Recent WAL measurements indicated stronger SOC in bubble-free devices, suggesting that fabrication methods that avoid bubble formation should be used \cite{Yang2017}. This result was supported by another recent work that compared two types of device: one with graphene on single-layer WS$_2$, and one with bulk WS$_2$ on graphene \cite{Wakamura2017}. As revealed by WAL measurements, the graphene/monolayer-WS$_2$ device exhibited a SOC strength at least one order of magnitude stronger than the bulk-WS$_2$/graphene device. This difference could be explained by the higher stiffness of the bulk WS$_2$, which would preclude a conformal fit to the graphene layer and lead to the formation of bubbles at the interface. Beyond PL, Raman spectroscopy can also be used to quantify the graphene/TMDC interaction \cite{Li2017}, but to date this has not been used in connection with spin transport measurements.

X-ray photoemission spectroscopy (XPS) is also a valuable tool for analyzing the quality of a graphene/TMDC heterostructure. In particular, XPS can be used to estimate the density of chalcogenide vacancies in the TMDC layer \cite{Avsar2014}, which is a common type of defect in these materials \cite{Qiu2013}. These vacancies can lead to strong localized SOC in the graphene layer, which could have a large impact on the spin transport properties. However, a detailed analysis of the role of these defects has yet to be undertaken.

Finally, it is important to understand the type of scattering induced in graphene by various impurities and defects. As we will show hereafter, the contribution of intervalley scattering has a fundamental impact on the nature of spin relaxation and spin current generation in graphene/TMDC heterostructures. The strength of intervalley scattering is typically deduced via weak localization (WL) measurements, but the presence of SOC makes it very difficult to disentangle the contributions of intervalley scattering and spin relaxation \cite{Mccann2006, Mccann2012}. Currently there is no alternative manner to extract the intervalley scattering rate, but such a technique would prove extremely useful for the analysis of graphene/TMDC devices.
%%%%%%%%%%%%%%%%%%%SECTION%%%%%%%%%%%%%%%%%
\section{Theory of graphene/TMDC heterostructures} \label{sec_theory}
%%%%%%%%%%%%%%%%%%%%%%%%%%%%%%%%%%%%%%%%
In the past few years, clear signatures of proximity-induced SOC have been measured in graphene/TMDC heterostructures \cite{Avsar2014, Wang2015, Wang2016, Yang2016, Yan2016, Yang2017, Ghiasi2017, Benitez2017, Omar2017, Omar2017b, Dankert2017, Volkl2017, Wakamura2017, Zihlmann2018, Friedman2018}. In parallel, these systems have been studied with {\it ab initio} and tight-binding methods \cite{Wang2015, Gmitra2016, Yang2016, Alsharari2016}, which yield quite consistent results for the SOC induced in graphene: a SOC strength on the order of 1 meV, the presence of Rashba SOC, and the appearance of a new type of SOC denoted valley-Zeeman SOC \cite{Yang2016}, which turns out to dominate over the other ones. These calculations also predicted that among all the TMDCs, WS$_2$, WSe$_2$, MoS$_2$, and MoSe$_2$ are the most suitable to be used as enabling substrates, given their imprint on the low-energy electronic properties of graphene and the fact that the graphene bands of interest lie within the band gap of the TMDC \cite{Gmitra2016}. 

%%%%%%%%%%%%%%%%%SUBSECTION%%%%%%%%%%%%%%%%%
\subsection{Electronic model} \label{sec_theory_model}
%%%%%%%%%%%%%%%%%%%%%%%%%%%%%%%%%%%%%%%%
To capture the main features of the {\it ab initio} calculations, a tight-binding (TB) model for the graphene layer was developed using group theory by Gmitra and coworkers \cite{Gmitra2016, Kochan2017}. In this model, all SOC terms allowed by symmetry are incorporated in the graphene Hamiltonian, and by fitting the energy bands and spin texture to the {\it ab initio} results, the strength of the various SOC parameters can be estimated. The spin texture is simply the spin polarization of an energy band as a function of the electron wave vector $\vec{k}$, which is related to the electron momentum by $\vec{p} = \hbar \vec{k}$. The model is given as
\begin{equation} \label{eq:first}
H=H_\text{orb}+H_\text{so},
\end{equation}
where the first term
\begin{equation}
H_\text{orb} = \hbar v_\text{F} ( \kappa \sigma_x k_x + \sigma_y k_y) + \Delta \sigma_z
\end{equation}
is the orbital part of electrons in monolayer graphene, representing the linear Dirac bands with $v_\text{F}$ the Fermi velocity. The hexagonal crystal structure of graphene can be viewed as two interpenetrating triangular sublattices, typically denoted A and B, and $\sigma_i$ are the Pauli matrices acting on this sublattice degree of freedom. The TMDC substrate creates an imbalance in the average electric potential felt by the A and B sublattices, which is given by $\Delta$. In the absence of SOC, this opens a band gap of 2$\Delta$ in the graphene bands. The band structure of graphene includes Dirac cones at both the K and K$'$ points, also called valleys, of reciprocal space. The valley index is given by $\kappa=1(-1)$ for the K (K$'$) valley. The second term,
\begin{equation} \label{eq:h_cont1}
H_\text{so} = H_\text{I} + H_\text{VZ} + H_\text{R} + H_{\Delta \text{PIA}} + H_\text{PIA},
\end{equation}
with
\begin{align} \label{eq:h_cont2} 
H_\text{I} &= \lambda_\text{I} \kappa \sigma_z s_z, \nonumber \\
H_\text{VZ} &= \lambda_\text{VZ} \kappa s_z, \nonumber \\
H_\text{R} &= \lambda_\text{R} (\kappa \sigma_x s_y - \sigma_y s_x), \nonumber \\
H_{\Delta \text{PIA}} &= a \lambda_{\Delta \text{PIA}} (k_x s_y - k_y s_x), \nonumber \\
H_\text{PIA} &= a \lambda_\text{PIA}  \sigma_z(k_x s_y - k_y s_x),
\end{align}
represents the proximity-induced enhancement of SOC in the graphene layer, where $s_i$ are the Pauli matrices for spin and $a = 2.46~\text{\AA}$ is the graphene lattice constant. $H_\text{I}$ is the intrinsic SOC in graphene, also known as Kane-Mele SOC, which opens a topological gap 2$\lambda_\text{I}$ at the Dirac point \cite{KaneMele2005}. $H_\text{VZ}$ is a valley-Zeeman term, which spin polarizes the bands out of the graphene plane with opposite orientation in the K and K$'$ valleys, also known as spin-valley locking. $H_\text{R}$ is a Rashba SOC with strength $\lambda_\text{R}$, arising from a perpendicular electric field. $H_{\Delta \text{PIA}}$ is a second-order Rashba term that causes a $k$-linear splitting of the bands, as in traditional 2D electron gases (2DEGs) with Rashba SOC \cite{Rashba1984}. Finally, $H_\text{PIA}$ is SOC that appears due to the absence of horizontal reflection symmetry and which renormalizes the Fermi velocity. Except for the PIA terms, this Hamiltonian is the same as that considered in other works \cite{Wang2015, Yang2016, Wang2016, Alsharari2016}. The values of these parameters are on the order of $\sim$1 meV except for the intrinsic SOC, which remains on the order of a few tens of $\upmu$eV, similar to pristine graphene. In all subsequent calculations of graphene/TMDC systems shown in this review, we use the parameters shown in Table \ref{table_params}, which are extracted from Ref.\ \citenum{Gmitra2016}. Here we would like to note that in addition to graphene in proximity to TMDCs, the Hamiltoninan in Eqs. (\ref{eq:first})-(\ref{eq:h_cont2}) is applicable to any hexagonal system in the presence of SOC, sublattice symmetry breaking, and a perpendicular electric field \cite{Kochan2017}.

%----------------------------------------TABLE-------------------------------------------%
\begin{table}[t]
\centering
%\begin{tabular}{c c c c c c c c}
%\toprule
%\multirow{2}{*}{TMDC} & $v_\text{F}$ & $\Delta$ & $\lambda_\text{I}$ & $\lambda_\text{VZ}$ & $\lambda_\text{R}$ & $\lambda_{\Delta \text{PIA}}$ & $\lambda_\text{PIA}$ \\
%& ($\mu$m/ps) & \multicolumn{6}{c}{(meV)} \\
%%& ($\mu$m/ps) & (meV) & ($\upmu$eV) & (meV) & (meV) & (meV) & (meV) \\
%%TMDC & $v_F$ & $\Delta$ & $\lambda_I$ & $\lambda_{VZ}$ & $\lambda_R$ & $\lambda_{\Delta PIA}$ & $\lambda_{PIA}$ \\
%\midrule
%MoS$_2$	& 0.85	& 0.52	& 0.025		& -0.26	& 0.13	& 0.51	& -1.73	\\
%MoSe$_2$	& 0.82	& 0.44	& -0.015		& -0.18	& 0.26	& -0.53	& 2.99	\\
%WS$_2$	& 0.85	& 1.31	& 0.095		& -1.12	& 0.36	& 1.42	& -2.4	\\
%WSe$_2$	& 0.82	& 0.54	& 0.030		& -1.19	& 0.56	& -0.08	& -2.62	\\
%\bottomrule
%\end{tabular}
\begin{tabular}{c c c c c}
\toprule
& MoS$_2$ & MoSe$_2$ & WS$_2$& WSe$_2$ \\
\midrule
$v_\text{F}$ ($10^6$ m/s)			& 0.85	& 0.82	& 0.85	& 0.82	\\
$\Delta$ (meV)					& 0.52	& 0.44	& 1.31	& 0.5		\\
$\lambda_\text{I}$ ($\upmu$eV)		& 25		& -15	& 95		& 30		\\
$\lambda_\text{VZ}$ (meV)		& -0.26	& -0.18	& -1.12	& -1.19	\\
$\lambda_\text{R}$ (meV)			& 0.13	& 0.26	& 0.36	& 0.56	\\
$\lambda_{\Delta \text{PIA}}$ (meV)	& 0.51	& -0.53	& 1.42	& -0.08	\\
$\lambda_\text{PIA}$ (meV)		& -1.73	& 2.99	& -2.4	& -2.62	\\
\bottomrule
\end{tabular}
\caption{Parameters for the model of graphene on a TMDC substrate. Reproduced from ref.\ \citenum{Gmitra2016}. with permission from the American Physical Society, \textcopyright 2016.}
\label{table_params}
\end{table}
%------------------------------------------------------------------------------------------%

%----------------------------------------FIGURE-----------------------------------------%
\begin{figure}[t]
\centering
\includegraphics[width=\columnwidth]{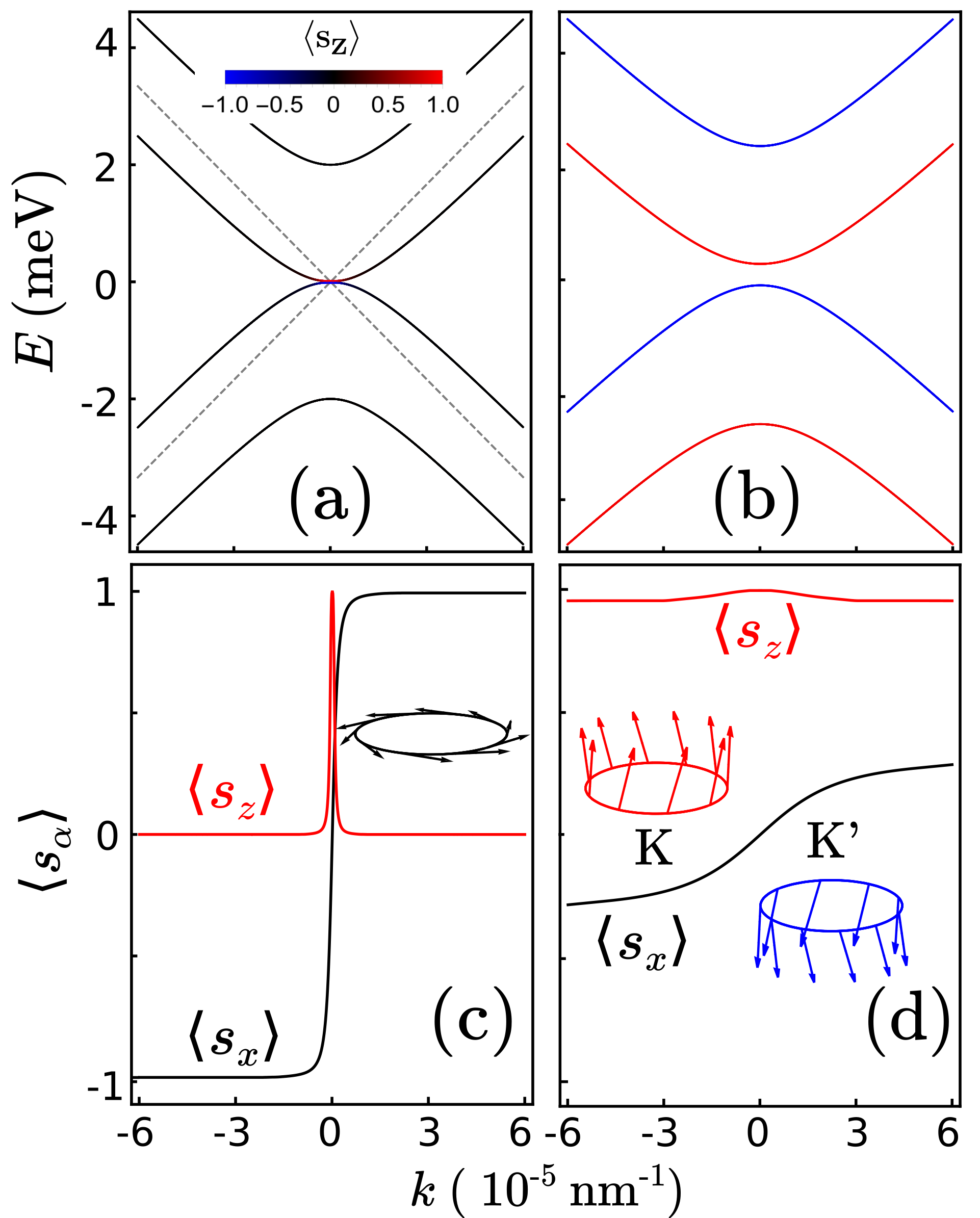}
\caption{Band structure of graphene (a) with only Rashba SOC ($\lambda_\text{R} = 1$ meV) and (b) in proximity to WS$_2$, where the color indicates the out-of-plane spin component. Dashed lines show the linear band structure of pristine single-layer graphene. The band structures are quite similar but due to the presence of valley-Zeeman SOC there is a noticeable change in the spin texture, which is shown in panel (c) for Rashba SOC and panel (d) for the WS$_2$ substrate. The insets show the spin texture at a given Fermi energy for both cases. Graphene/WS$_2$ model parameters are shown in Table \ref{table_params}.}
\label{fig:rashba-vs-vz}
\end{figure}
%------------------------------------------------------------------------------------------%

In Fig.\ \ref{fig:rashba-vs-vz}, we show a comparison between graphene with only Rashba SOC ($\lambda_\text{R} = 1$ meV), and graphene in proximity to WS$_2$. Both cases produce very similar energy bands, as seen in Figs.\ \ref{fig:rashba-vs-vz}(a,b), with a spin splitting of $\sim$2 meV and nonlinearity near the graphene Dirac point. However, the resulting spin textures are very different. For pure Rashba SOC, Fig.\ \ref{fig:rashba-vs-vz}(c), the splitting is accompanied by an in-plane spin texture associated with spin-momentum locking, where the spin is perpendicular to the momentum direction. For the graphene/TMDC case, Fig.\ \ref{fig:rashba-vs-vz}(d), the valley-Zeeman term induces an effective out-of-plane magnetic field which tilts the spin texture in the perpendicular direction, with opposite direction in each valley. This difference will have major implications for the spin relaxation, weak antilocalization, and the spin Hall effect, as we will show in the following sections. This result also highlights the importance of considering the spin texture when fitting to {\it ab initio} data. 

Finally, it is important to note that this model considers {\it only} the electronic structure of the graphene layer, with the pristine graphene band structure enhanced by extra terms that describe the proximity of the TMDC layer. This approach is justified because DFT simulations show that the graphene Dirac bands, including the Dirac point, sit within the band gap of the TMDC substrate \cite{Gmitra2016}; see for example Fig.\ \ref{fig:hanle-abs}(b). Therefore, in experiments charge transport occurs solely within the graphene layer, except when a sufficiently large gate voltage allows parallel transport to also occur in the TMDC layer \cite{Wang2015, Wang2016, Yang2017, Benitez2017, Dankert2017, Yan2016}.

%%%%%%%%%%%%%%%%%SUBSECTION%%%%%%%%%%%%%%%%%
\subsection{Microscopic theory of spin relaxation} \label{sec_theory_relaxation}
%%%%%%%%%%%%%%%%%%%%SECTION%%%%%%%%%%%%%%%%%
When electrons propagate in a solid, the typical approximation is to consider the spin as an inactive internal degree of freedom. However, in the presence of spin-orbit coupling this is not the case, as the coupling of spin with momentum implies that the spin can now play a role in transport and be affected by scattering. The actual spin dynamics will depend on the disorder profile and the nature of the spin-orbit coupling. In graphene/TMDC heterostructures, experimental results suggest that the relaxation mechanism occurs in the DP regime \cite{Wang2015, Yang2016, Yang2017, Volkl2017, Friedman2018}, where the spin precession of the electrons is interrupted by scattering events, inducing motional narrowing and yielding $\tau_\text{s} \propto 1/\tau_\text{p}$.

In DP theory, the electronic bands are split by the effect of the SOC field, in the same way that spin up and down electronic states split under the application of an external magnetic field. The model above yields wave functions in the sublattice and spin basis, $[A_\uparrow~B_\uparrow~A_\downarrow~B_\downarrow]^\text{T}$, and therefore does not provide clear information about the nature of this splitting. Thus, we use a downfolding procedure \cite{Lowdin1951, Shanavas2014, McCann2013} to transform the model given in Section \ref{sec_theory_model} into a more transparent form. This is done by expressing $H$ in the basis of the eigenstates of $H_\text{orb}$ followed by a projection onto the conduction and valence bands. At Fermi energies away from the Dirac point, this gives
\begin{equation} \label{eq:h_cont3}
H = H_\text{orb} + \frac{1}{2} \hbar \vec{\omega}(t) \cdot \vec{s},
\end{equation}
where the second term represents an effective magnetic field induced by the presence of SOC. The direction and strength of this effective magnetic field are contained in the spin precession frequency $\vec{\omega}$, whose components are
\begin{align}
&\hbar \omega_x (t) = -2(ak\lambda_{\Delta \text{PIA}} \pm \lambda_\text{R}) \sin (\theta(t)), \nonumber\\ 
&\hbar \omega_y (t) =  2(ak\lambda_{\Delta \text{PIA}} \pm \lambda_\text{R}) \cos (\theta(t)), \\ 
&\hbar \omega_z (t) = 2 \lambda_\text{VZ} \kappa(t), \nonumber
\end{align}
where $k$ is the wave vector magnitude and $\theta$ is the direction of $k$ with respect to $k_x$. The in-plane components of $\vec{\omega}$ give a Rashba-like spin texture, where the spin polarization is perpendicular to the direction of momentum. The $+(-)$ is for the conduction (valence) band, indicating that the PIA SOC enhances (reduces) the strength of the Rashba SOC. The out-of-plane component of $\vec{\omega}$ is determined by $\lambda_\text{VZ}$ and changes sign between valleys. Note that the intrinsic SOC does not appear; it has not been eliminated because of its small magnitude but rather because it plays no role in the spin texture.

Owing to the presence of charge scattering, each component of $\vec{\omega}$ will fluctuate in time, which is represented by the time dependence of $\theta$ and $\kappa$. If this fluctuation is uncorrelated, then the electron will not have any memory of its previous magnetic field, leading to a randomization of its spin. Because they depend only on $\theta$, the in-plane components of $\vec{\omega}$ are randomized according to the momentum relaxation time $\tau_\text{p}$. However, because $\omega_z$ depends only on the valley $\kappa$, this component is randomized in accordance with the intervalley scattering time $\tau_\text{iv}$. Based on these considerations, and assuming the DP regime, the relaxation rates of the out-of-plane and in-plane spins are \cite{Cummings2017}
\begin{align} \label{eq:rates_intra}
\left(\tau_\text{s}^{\perp}\right)^{-1} &= \left(2\frac{ak\lambda_{\Delta \text{PIA}} \pm \lambda_\text{R} }{\hbar}\right)^2 \tau_\text{p},\nonumber \\
\left(\tau_\text{s}^{\parallel}\right)^{-1} &= \left(2\frac{\lambda_\text{VZ} }{\hbar}\right)^2 \tau_\text{iv} +\frac{1}{2} \left(\tau_\text{s}^{\perp}\right)^{-1}.
\end{align}
The out-of-plane spin thus follows a typical Rashba-induced relaxation process, with an electron-hole asymmetric behavior that originates from the PIA SOC. However, the in-plane spin includes contributions from both the Rashba SOC and the valley-Zeeman SOC through the intervalley scattering time $\tau_\text{iv}$. Because in general $\lambda_\text{VZ} > \lambda_\text{R}$ and by definition $\tau_\text{iv} \ge \tau_\text{p}$, the in-plane spin relaxation should be dominated by valley-Zeeman SOC and intervalley scattering, causing the in-plane spins to relax much faster than the out-of-plane spins. Dividing the spin relaxation rates in Eq.\ (\ref{eq:rates_intra}) gives the expression for the spin lifetime anisotropy in graphene/TMDC systems,
\begin{equation} \label{eq:anisotropy}
\zeta \equiv \frac{\tau_\text{s}^\perp}{\tau_\text{s}^\parallel} = \left( \frac{\lambda_\text{VZ}}{ak\lambda_{\Delta \text{PIA}} \pm \lambda_\text{R}} \right)^2
\left(\frac{\tau_\text{iv}}{\tau_\text{p}}\right)+\frac{1}{2}.
\end{equation}
Using DFT values for graphene on WSe$_2$ \cite{Gmitra2016}, and assuming relatively strong intervalley scattering ($\tau_\text{iv} \approx 5 \tau_\text{p}$), we obtain a spin lifetime anisotropy of $\zeta \approx 20$. This represents a qualitatively different regime of spin relaxation than the usual case of 2D Rashba systems, where without valley Zeeman SOC the anisotropy is 1/2, with the in-plane spins relaxing more slowly than the out-of-plane spins. This is also different from other proposed spin relaxation mechanisms, such as the ferromagnetic contacts, magnetic impurities, or spin-pseudospin coupling, which all give isotropic spin relaxation. Meanwhile, strain-induced gauge fields are predicted to yield large anisotropy, but the spin lifetimes are orders of magnitude larger than what is expected in graphene/TMDC systems. To date, measurements of spin lifetime anisotropy in  graphene on SiO$_2$ or hBN substrates have all yielded $\zeta \approx 1$ \cite{Raes2016, Raes2017, Guimaraes2014, Tombros2008, Ringer2017}. Thus, a giant anisotropy should be an experimental fingerprint of SOC proximity effects induced in graphene by TMDCs.

%----------------------------------------FIGURE-----------------------------------------%
\begin{figure}[t]
\centering
\includegraphics{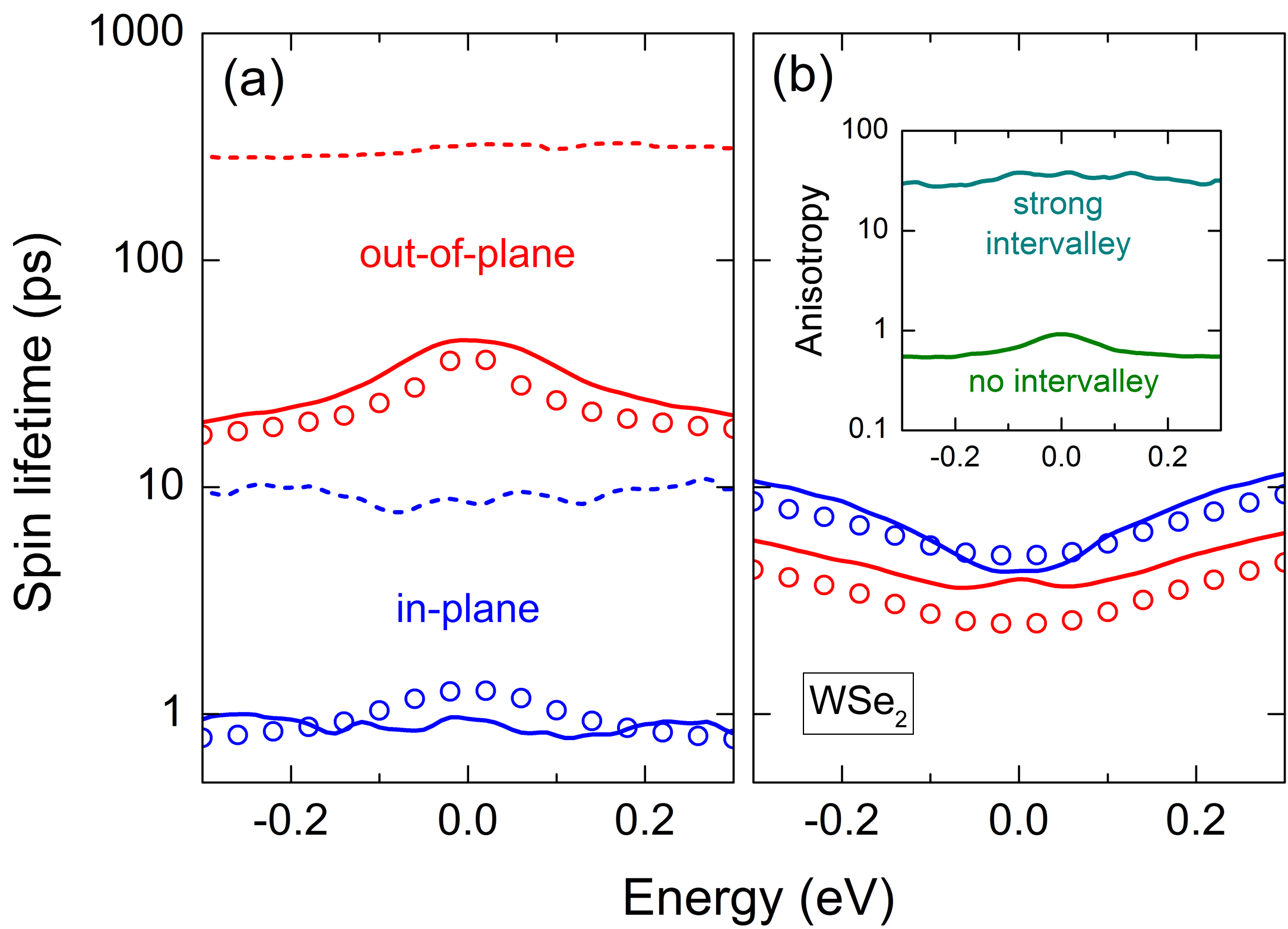}
\caption{Numerical simulations of the in-plane (blue) and out-of-plane (red) spin lifetimes in graphene/WSe$_2$ for (a) strong intervalley scattering ($U_\text{p}=2.8$ eV, $n_\text{p} = 0.1\%$, and $\xi = \sqrt{3}a$), and (b) weak intervalley scattering ($U_\text{p}=0.5$ eV, $n_\text{p} = 1\%$, and $\xi = \sqrt{3}a$). Dashed lines in (a) show the case for $n_\text{p} = 1\%$. In the inset we show the spin lifetime anisotropy for these two cases. Graphene/WSe$_2$ model parameters are shown in Table \ref{table_params}\,. Figure reproduced from ref.\ \citenum{Cummings2017} with permission from the American Physical Society, \textcopyright 2017.}
\label{fig:micro-anisotropy}
\end{figure}
%-----------------------------------------------------------------------------------------%

Equation (\ref{eq:anisotropy}) shows that the role of the PIA SOC is to induce an electron-hole asymmetry in the spin lifetime anisotropy, as it enhances (cancels) the Rashba SOC in the conduction (valence) band. The degree of this asymmetry depends on the magnitude of $\lambda_{\Delta \text{PIA}}$ and on the range of carrier densities $n$ that are probed experimentally, as $k \approx \sqrt{\pi n}$. Taking the values from Table \ref{table_params} and assuming a typical experimental carrier density range of $n \in [-3,3] \times 10^{12}$ cm$^{-2}$, $\zeta$ is expected to vary by a factor of 2 to 4 for graphene on MoS$_2$, MoSe$_2$, or WS$_2$, while for WSe$_2$ it would only vary by $\sim$5\%. To date, measurements of the spin lifetime anisotropy in graphene/TMDC heterostructures have reported little to no monotonic dependence of $\zeta$ in this carrier density range, suggesting that either the PIA SOC is small or that other relaxation mechanisms could be masking the electron-hole asymmetry \cite{Ghiasi2017, Benitez2017}.

To validate Eqs. (\ref{eq:rates_intra}) and (\ref{eq:anisotropy}), in Fig.\ \ref{fig:micro-anisotropy} we present a numerical simulation of spin lifetimes in graphene on WSe$_2$. The simulations were performed considering electron-hole puddle disorder, which consists of localized regions of positive or negative charge density distributed throughout the graphene lattice. These regions, or puddles, of charge typically arise from charged dopants in the graphene device, and can be modeled as a random distribution of Gaussian-shaped variations of the local electrostatic potential \cite{Martin2007, Deshpande2009, Zhang2009b, Adam2009}. The strength of this disorder profile is characterized by the puddle height $U_\text{p}$, the puddle concentration $n_\text{p}$, and the puddle width $\xi$, and by tuning these parameters one can tune the magnitude of $\tau_\text{p}$ as well as the ratio $\tau_\text{iv}/\tau_\text{p}$. For $U_\text{p}=2.8$ eV, $n_\text{p} = 0.1\%$, and $\xi = \sqrt{3}a$ these puddles produce strong intervalley scattering \cite{Zhang2009}, which according to our theory should induce giant anisotropy. This is indeed the case, as shown in Fig.\ \ref{fig:micro-anisotropy}(a), where we find $\tau_\text{s}^{\perp} = 20\text{--}50$ ps and $\tau_\text{s}^{\parallel} \approx 1$ ps. The open circles are the values of $\tau_\text{s}$ estimated from Eq.\ (\ref{eq:rates_intra}), showing good agreement between the numerical simulations and the spin dynamics model. As shown by the dashed lines, increasing the puddle density to $n_\text{p} = 1\%$ scales $\tau_\text{s}$ by a factor of 10, showing that the inverse relationship between $\tau_\text{s}$ and $\tau_\text{p,iv}$ holds.

The in-plane spin relaxation rate in Eq.\ (\ref{eq:rates_intra}) was derived assuming the motional narrowing regime, when intervalley scattering is strong. When intervalley scattering is weak this assumption no longer applies and the first term in the expression for $1 / \tau_\text{s}^\parallel$ disappears, leading to $\zeta = 1/2$. Figure \ref{fig:micro-anisotropy}(b) shows the case for weak intervalley scattering, with $U_\text{p}=0.5$ eV, $n_\text{p} = 1\%$, and $\xi = \sqrt{3}a$, where we see that the anisotropy indeed collapses toward 1/2. The inset of Fig.\ \ref{fig:micro-anisotropy}(b) shows the comparison between the anisotropy for weak and strong intervalley scattering. These results highlight the strong connection between intervalley scattering and the in-plane spin lifetime in graphene/TMDC heterostructures.

As shown above, the nature of spin relaxation in graphene/TMDC heterostructures depends strongly on the presence or absence of intervalley scattering. In general, intervalley scattering is caused by defects whose size is on the order of the graphene lattice constant. This includes structural defects such as dislocations, grain boundaries, vacancies, etc. \cite{Cao2010, Nemes2013}, as well as chemical adsorbates that could be deposited during device fabrication \cite{Ni2010}. TMDCs are known to suffer from chalcogenide vacancies \cite{Qiu2013, Avsar2014}; these might also induce short-range Coulomb potentials in graphene, leading to intervalley scattering. Measuring $\tau_\text{p}$ in these systems is straightforward, as it can be deduced from knowing the mobility and charge density, assuming diffusive transport. Determining $\tau_\text{iv}$ requires a measurement of weak localization \cite{Mccann2006}. However, in graphene/TMDC systems the strong SOC leads to weak antilocalization, which does not allow for a straightforward extraction of $\tau_\text{iv}$ \cite{Mccann2012}. So far, the best that has been done for a graphene/TMDC system is to measure WL in a region of the device that is not covered by the TMDC, and to assume that value as an upper bound of $\tau_\text{iv}$ in the graphene/TMDC region \cite{Yang2016}. Measurements of WL in graphene systems yield $\tau_\text{iv} / \tau_\text{p}$ in the range of 3 to 20 depending on the sample quality and Fermi energy \cite{Wu2007, Ki2008, Yang2016}. In general, $\tau_\text{p}$ is in the range of tens of fs and $\tau_\text{iv}$ is on the order of hundreds of fs to a few ps.

In this section, we presented a theoretical framework for spin relaxation in graphene/TMDC heterostructures, and showed that spin lifetime anisotropy is an undeniable fingerprint of proximity-induced SOC in graphene, provided that intervalley scattering is present. However, in real samples and devices, different sources of spin relaxation not related to the TMDC may also be present, including the ferromagnetic contacts, magnetic impurities, spin-pseudospin coupling, etc. In order to determine whether the TMDC is impacting spin relaxation in the graphene layer, one should measure the spin relaxation anisotropy, since all of these other mechanisms yield isotropic spin relaxation. The typical Hanle setup used to measure spin relaxation in graphene only probes the in-plane spin relaxation and gives zero information about the out-of-plane component. Therefore, in the following section we will review a generalized Hanle measurement that allows for the detection of both the in-plane and out-of-plane spin relaxation behavior.

%%%%%%%%%%%%%%%%%%%%SECTION%%%%%%%%%%%%%%%%%
\section{Lateral spin valves and Hanle precession in graphene/TMDC heterostructures} \label{spin-precession}
%%%%%%%%%%%%%%%%%%%%%%%%%%%%%%%%%%%%%%%%%

\begin{figure}[t]
\centering
\includegraphics{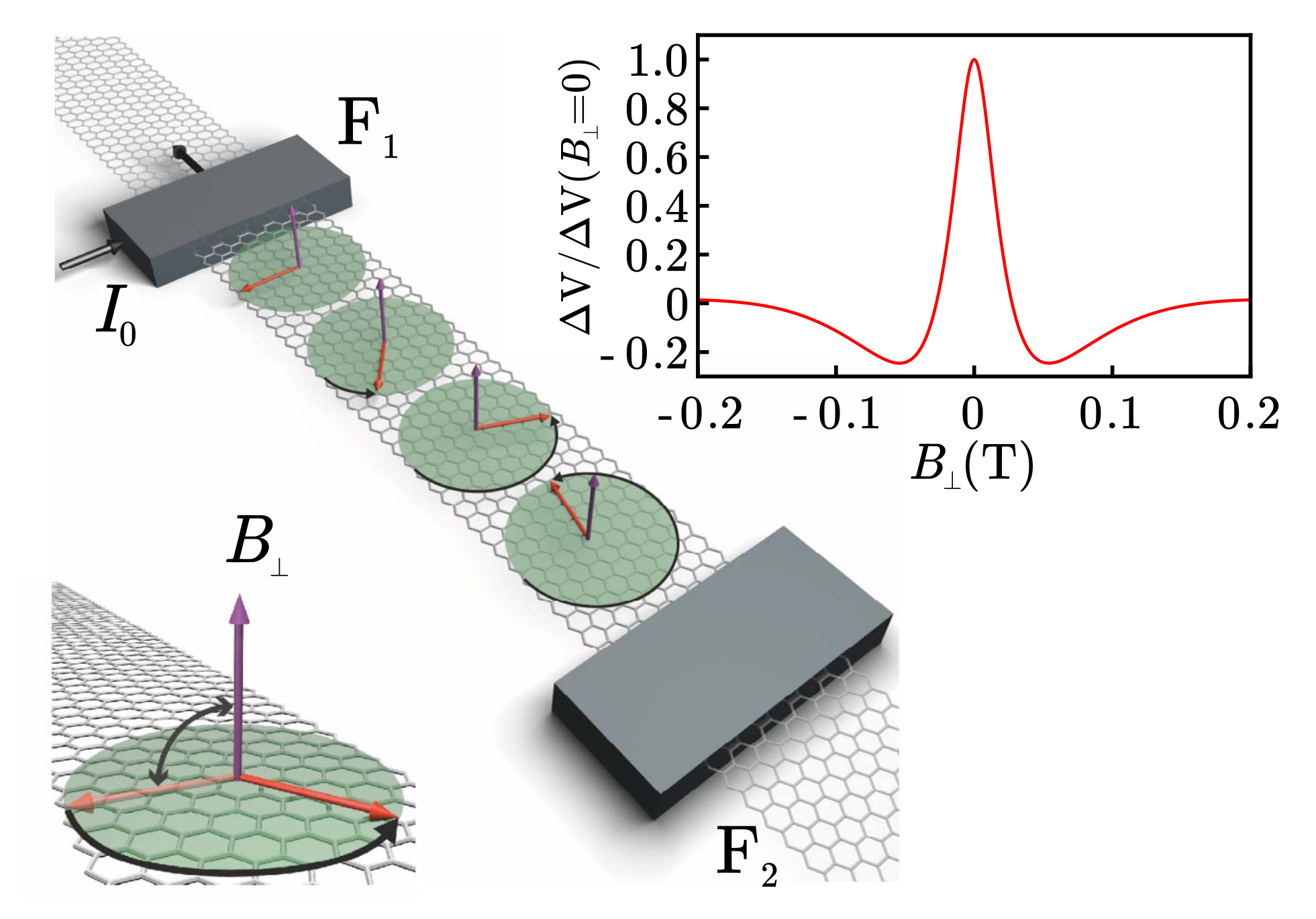}
\caption{Schematic of a LSV in the presence of a perpendicular magnetic field $B$. At the injector a spin-polarized charge current is injected, leading to the formation of spin accumulation below F$_1$ which propagates diffusively through the channel. The out-of-plane magnetic field drives spin precession around the perpendicular axis, giving rise to dephasing and suppression of the nonlocal voltage, as shown in the inset where we show the voltage difference normalized by its value at zero magnetic field as function of $B$ . Modified from ref.\citenum{Raes2016} with permission from Springer Nature \textcopyright 2016. }
\label{fig:typical-hanle}
\end{figure}

The nonlocal lateral spin valve (LSV) is by far the most common technique for measuring the spin lifetime in graphene. The concept behind the LSV dates back to 1985, when Johnson and Silsbee proposed a method \cite{Johnson1985} to induce a nonequilibrium spin density in a nonmagnetic material by injecting a spin-polarized charge current. The typical nonlocal LSV experiment is sketched in Fig.\ \ref{fig:typical-hanle}, where two ferromagnets F$_1$  and F$_2$ are placed on top of a nonmagnetic metal and a spin-polarized current $I_0$ is forced to flow from the injector at F$_1$ to the left of the device. This leads to the formation of a spin density accumulation $\bm{n}_\text{s}$ in the region below the injector, which points parallel to the magnetization direction of F$_1$. The electrically-induced spin density then propagates diffusively through the material and can be detected by F$_2$, allowing for the determination of $n_\text{s}$, $\lambda_\text{s}$, and $\tau_\text{s}$.

In the presence of a magnetic field $B$, the dynamics of the spin density can be described by the Bloch-Torrey equations (BTEs)
\begin{equation} \label{eq:BTE}
D_\text{s} \nabla^2 \bm{n}_\text{s}+\frac{\partial \bm{n}_\text{s}}{\partial t}=- \Omega\cdot \bm{n}_\text{s}+\omega_B \bm{n}_\text{s}\times \bm{\hat{e} }_{B},
\end{equation} 
where $\omega_B$ is the Larmor precession frequency induced by the magnetic field, $D_\text{s}$ is the spin diffusion coefficient, $\bm{\hat{e}}_B$ is a unit vector pointing parallel to the magnetic field, and $\Omega$ is a matrix representing the spin relaxation rates. In the coordinate basis $\{\bm{\hat{e}}_x,\bm{\hat{e}}_\parallel,\bm{\hat{e}}_\perp\}$, where the unit vectors point in the propagation, magnetization, and out-of plane directions respectively, the relaxation rate matrix has the form
\begin{equation}	
\Omega = 
\left[\begin{array}{ccc}
1/\tau_\text{s}^{\parallel}& 0 &0\\
0&1/\tau_\text{s}^{\parallel} &0\\
0&0&1/\tau_\text{s}^{\perp}
\end{array} \right].
\end{equation}

In the absence of magnetic field, the BTEs are decoupled and the spin density parallel to the injector's magnetization decays exponentially along the length of the channel. When the spin density arrives at the detector, it is transformed into a voltage that will depend on the relative orientation between the magnetization of $F_1$ and $F_2$. To eliminate background effects, the voltage is measured with the injector/detector magnetized in parallel (P) and antiparallel (AP) orientations. The voltage difference can be expressed as 
\begin{equation} \label{differentialresistance}
\Delta V = P e D_\text{s} n_\text{s}^{\parallel}(L) = P^2 \frac{I_0}{\sigma} \frac{\lambda_\text{s}^\parallel}{w} \text{e}^{-L/\lambda_\text{s}^\parallel},
\end{equation}
where $P$ is the spin injection/detection efficiency, $w$ is the width of the channel, $\sigma$ is the conductivity, $\lambda_\text{s}^\parallel = \sqrt{D_\text{s} \tau_\text{s}^\parallel}$ is the relaxation length of spins polarized in the graphene plane, and $L$ is the distance between F$_1$ and F$_2$. This measurement thus provides a direct probe of the in-plane spin relaxation length. In order to also extract the spin relaxation times, an external magnetic field must be applied, as discussed in the next sections.
%----------------------------------------FIGURE-----------------------------------------%
%-----------------------------------------------------------------------------------------%

%%%%%%%%%%%%%%%%%%%SUBSECTION%%%%%%%%%%%%%%%%%
\subsection{Out-of-plane magnetic field} \label{sec_hanle_out}
%%%%%%%%%%%%%%%%%%%%%%%%%%%%%%%%%%%%%%%%%%

In the presence of a perpendicular magnetic field, the spin will precess in the plane of the material, as shown in Fig.\ \ref{fig:typical-hanle}. In this situation, the out-of-plane component of the spin density decouples from the BTEs, leading to two coupled equations that depend on the in-plane relaxation time. Their solution yields an expression for the nonlocal voltage that is quite similar to Eq.\ (\ref{differentialresistance}),
\begin{equation} \label{eq:final-hanlez}
\Delta V(B) = \text{Re}\left\{P^2 \frac{I_0}{\sigma} \frac{\tilde{\lambda}_\text{s}^\parallel}{w} \text{e}^{-L/\tilde{\lambda}_\text{s}^{\parallel}} \right\},
\end{equation} 
where $\tilde{\lambda}_\text{s}^{\parallel}$ is a complex renormalization of the spin relaxation length due to the effect of the magnetic field,
\begin{equation}
\tilde{\lambda}_\text{s}^{\parallel} = \frac{\lambda_\text{s}^\parallel}{\sqrt{1+i \omega_B \tau_\text{s}^{\parallel}}}.
\end{equation}
The applied magnetic field can therefore modulate and completely suppress the spin signal, as shown in the inset of Fig.\ \ref{fig:typical-hanle}. This modulation allows the extraction of the in-plane relaxation time $\tau_\text{s}^\parallel$ and the spin diffusion coeffcient $D_\text{s}$ by fitting the measured $\Delta V (B)$ to Eq.\ (\ref{eq:final-hanlez}). However, this measurement does not provide information about the out-of-plane component of the spin.

%%%%%%%%%%%%%%%%%%%SUBSECTION%%%%%%%%%%%%%%%%%
\subsection{In-plane magnetic field} \label{sec_hanle_inplane}
\begin{figure}[!h]
\includegraphics[width=\columnwidth]{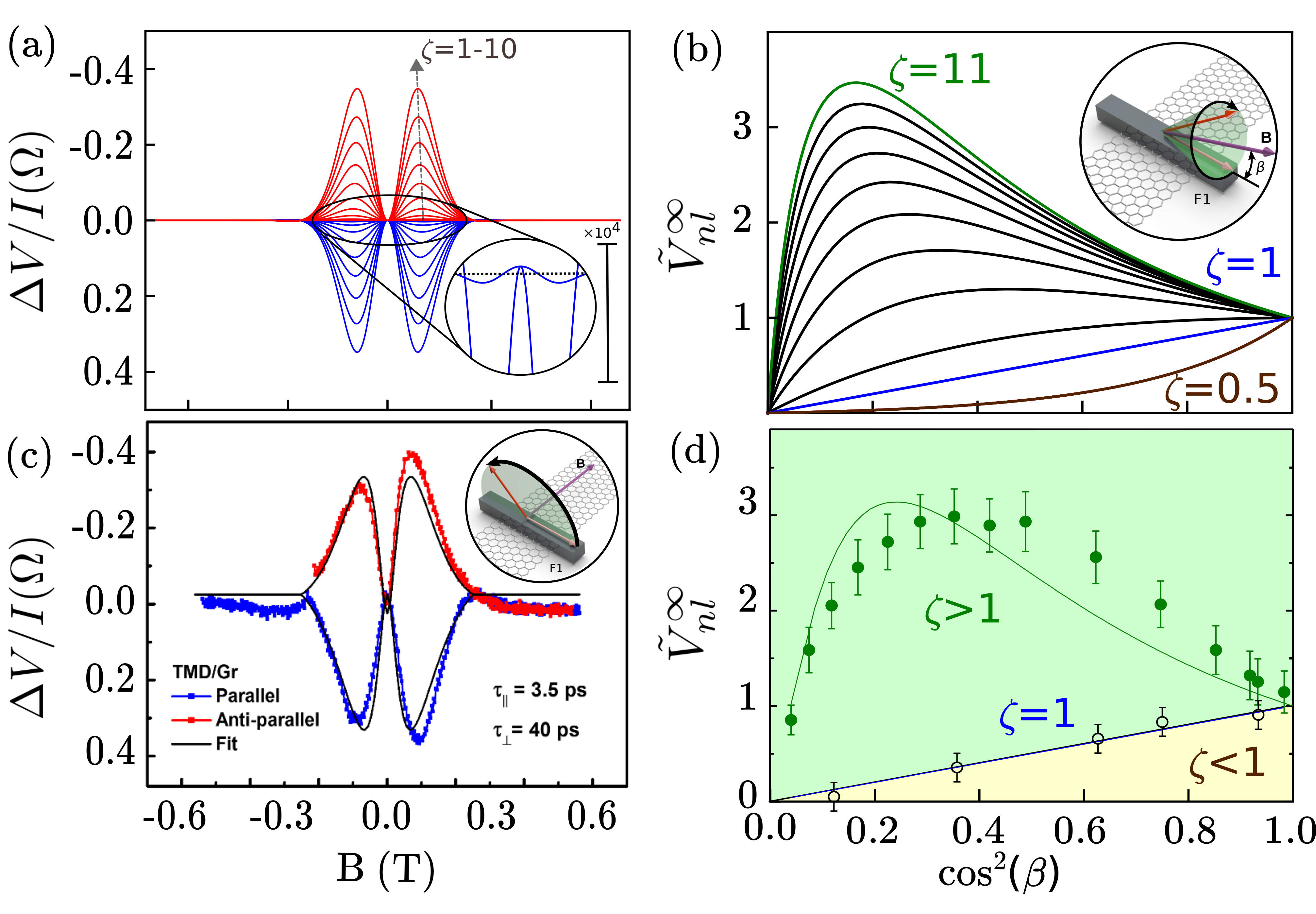}
\caption{(a) Theoretical nonlocal spin signal vs. in-plane magnetic field for different values of spin lifetime anisotropy. The inset shows a zoom-in of the cases, $\zeta = 1$ and $\zeta = 2$ for the parallel configuration, the dashed line represent the zero of the $y$-axis. (b) Normalized high-field nonlocal voltage as a function of the oblique magnetic field angle $\beta$, for different anisotropy values. Curves in red, blue, and green highlight the cases $\zeta = 11$, $1$, and $1/2$ respectively. The inset shows a schematic of the experimental setup. (c) Experimental values of the nonlocal resistance in graphene/MoSe$_2$, modulated by an in-plane magnetic field, reproduced from ref.\ \citenum{Ghiasi2017} with permission from the American Chemical Society \textcopyright 2017. {The inset shows a schematic of the experimental setup.} (d) Normalized high-field nonlocal voltage as a function of the oblique angle $\beta$ for graphene/SiO$_2$ and graphene/WS$_2$, reproduced from ref.\ \citenum{Benitez2017} with permission from Springer Nature \textcopyright 2017. In both (c) and (d) there is a tenfold enhancement of the anisotropy compared to graphene on SiO$_2$. }
\label{fig:hanle-experiments}
\end{figure}

%%%%%%%%%%%%%%%%%%%%%%%%%%%%%%%%%%%%%%%%%%%
In order probe the out-of-plane spin relaxation, a magnetic field can be applied parallel to the propagation direction $\bm{\hat{e}_x}$, {as shown in the inset of Fig.\ \ref{fig:hanle-experiments}(c),} which allows the spin to precess out of the graphene plane. For this case, the equation for $n_\text{s}^x$ decouples from the others, leaving a system of two equations that this time involve $\tau_\text{s}^{\parallel}$ and $\tau_\text{s}^{\perp}$. These equations can be solved by rewriting the BTEs in matrix form,
\begin{equation}\left[\begin{array}{c}
n_\text{s}^{\parallel}(x)\\
n_\text{s}^{\perp}(x)\\
\end{array}\right] = \text{e}^{-x \Lambda^{-1}} \left[\begin{array}{c}
n_\text{s}^{\parallel}(0)\\
n_\text{s}^{\perp}(0)\\
\end{array}\right],
\end{equation}
where
\begin{align}
\Lambda^{-2}={\left(\lambda_\text{s}^\parallel \right)^{-2}}\left[
\begin{array}{cc}
1&-\omega_B \tau_\text{s}^\parallel\\
\omega_B \tau_\text{s}^\parallel&1/\zeta
\end{array}
\right]
\end{align} 
is the matrix generalization of $\lambda_\text{s}^{-2}$. The problem thus reduces to a computation of the matrix $\exp(-x\Lambda^{-1})$. The solution is too cumbersome to show here; instead we present the results for some typical experimental values in Fig.\ \ref{fig:hanle-experiments}(a) with a range of anisotropy values. {When compared to the Hanle curve with $\zeta = 1$ shown in the inset, one can immediately see the impact of a large spin lifetime anisotropy.} At zero magnetic field, the nonlocal spin signal is small because of the small in-plane spin lifetime. At finite magnetic fields, the spin will experience a combination of the in-plane and out-of-plane spin relaxation times. For spin lifetime anisotropy $\zeta > 1$, the effective spin lifetime will thus be enhanced compared to the pure in-plane lifetime, leading to an enhanced nonlocal signal at finite magnetic fields. This can be clearly seen in Fig.\ \ref{fig:hanle-experiments}(a), where the maximum of the nonlocal signal increases with increasing $\zeta$, and is in contrast to the isotropic case in the inset, where the nonlocal signal is maximized at $B=0$.

%----------------------------------------FIGURE-----------------------------------------%
%------------------------------------------------------------------------------------------%

%%%%%%%%%%%%%%%%%%%SUBSECTION%%%%%%%%%%%%%%%%%
\subsection{Oblique magnetic field} \label{sec_hanle_oblique}
%%%%%%%%%%%%%%%%%%%%%%%%%%%%%%%%%%%%%%%%%%

An alternative approach for measuring the spin lifetime anisotropy is to align the magnetic field perpendicular to the transport direction, in the $\bm{\hat{e}}_\parallel\text{--}\bm{\hat{e}}_\perp$ plane, as shown in the inset of Fig.\ \ref{fig:hanle-experiments}(b). In the limit of large magnetic field, the components of spin perpendicular to $B$ will become completely dephased, and the detector will only measure the component parallel to the magnetic field, given by
\begin{equation} \label{spin_parallel_b}
n_\text{s}^{\beta} \equiv \bm{n}_\text{s} \cdot \bm{\hat{e} }_{B}=|\bm{n}_\text{s}|\cos(\beta),
\end{equation}
 where $\beta$ is the angle between the field and $\bm{\hat{e}}_\parallel$. At finite $\beta$, the effective spin relaxation time $\tau_\text{s}^\beta$ is a combination of the in-plane and out-of-plane spin relaxation \cite{Raes2016, Raes2017}, 
\begin{equation} \label{length_aniso}
\tau_\text{s}^\beta = \tau_\text{s}^{\parallel} \left(\cos^2(\beta) + \frac{1}{\zeta}\sin^2(\beta)\right)^{-1} \equiv \tau_\text{s}^{\parallel} f(\beta,\zeta)^{-1}.
\end{equation}
Equations (\ref{spin_parallel_b}) and (\ref{length_aniso}) enable the extraction of the spin anisotropy by measuring the nonequilibrium spin density as a function of $\beta$. Experimentally, the nonlocal voltage at large magnetic field is normalized by the value at $B=0$, and has the form
\begin{equation}
\tilde{V}_\text{nl}^{\infty}(\beta,\zeta) = \sqrt{f(\beta,\zeta)} \text{exp}\left[-\frac{L}{\lambda_\text{s}^\parallel}\left(\frac{1}{\sqrt{f(\beta,\zeta)}} -1   \right)\right]\cos^2(\beta).
\end{equation}
This expression reduces to $\cos^2(\beta)$ when the spin relaxation is isotropic. In Fig.\ \ref{fig:hanle-experiments}(b), we show $\tilde{V}_\text{nl}^{\infty}(\beta,\zeta)$ as a function of $\cos^2(\beta)$ for different values of anisotropy. For $\zeta < 1$ $(\zeta > 1)$, the curve falls below (above) the curve for $\zeta = 1$.

%%%%%%%%%%%%%%%%%%%SUBSECTION%%%%%%%%%%%%%%%%%
\subsection{Hanle measurements in graphene/TMDC heterostructures} \label{sec_hanle_exp}
%%%%%%%%%%%%%%%%%%%%%%%%%%%%%%%%%%%%%%%%%%

\begin{figure}[t]
\centering
\includegraphics[width=\columnwidth]{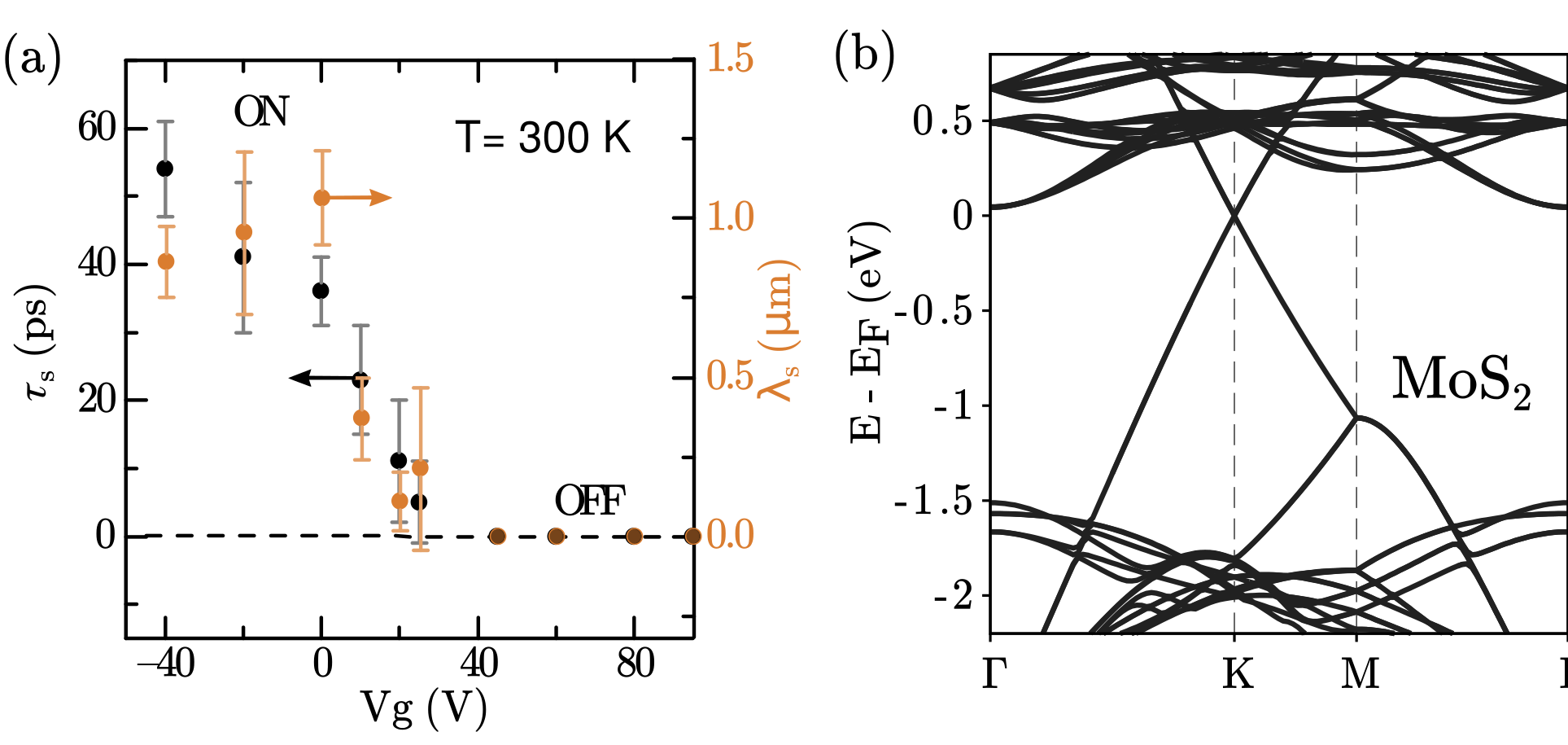}
\caption{(a) Spin relaxation times and lengths obtained for a graphene/MoS$_2$ heterostructure, reproduced from ref.\ \citenum{Dankert2017} with permission from Springer Nature, \copyright 2017. (b) Band structure for graphene/MoS$_2$ reproduced from ref.\ \citenum{Gmitra2016} with permission from the American Physical Society, \copyright 2016. For this heterostructure, the graphene Dirac cone lies close to the conduction band of MoS$_2$, allowing for transport within its bulk bands for positive gate voltage. Given the strong SOC in the TMDC, the spin relaxes immediately after its absorption, leading to the switching behavior.}
\label{fig:hanle-abs}
\end{figure}

In the first experimental measurement of Hanle precession in graphene/TMDC systems, the nonlocal spin signal of a LSV in a graphene/MoS$_2$ heterostructure was shown to be tunable with an electrical gate, exhibiting an on/off behavior that persisted up to 200 K \cite{Yan2016}. A subsequent experiment reproduced these results at room temperature, also in graphene/MoS$_2$ \cite{Dankert2017}. Figure \ref{fig:hanle-abs}(a) shows this on/off behavior for the room temperature measurements, with the spin relaxation time dropping to zero at positive gate voltages. Although it is tempting to associate this behavior with proximity-induced spin-orbit coupling, it is in fact a spin absorption effect. As shown in Fig.\ \ref{fig:hanle-abs}(b), the electronic structure of graphene/MoS$_2$ has its Fermi level close to the MoS$_2$ conduction band, meaning that a positive gate voltage will push the Fermi level into the TMDC conduction band and induce parallel spin conduction in both graphene and MoS$_2$. The latter, having SOC of the order of hundreds of meV \cite{Liu2013}, produces a much faster spin relaxation than in graphene and effectively kills the spin signal. On the other hand, below this threshold voltage spins are only transported in graphene bands, where longer $\tau_\text{s}^\parallel$ and $\lambda_\text{s}^\parallel$ are expected, and a spin signal can be measured. This theory of spin absorption has also been supported by measurements of the Schottky barrier between graphene and MoS$_2$ \cite{Dankert2017}.

While the suppression of spin signal at positive gate voltages is attributable to spin absorption, the magnitude of $\tau_\text{s}^\parallel$ at negative gate voltages, with values less than 60 ps, suggests a proximity-induced SOC in the graphene layer. This is in contrast to graphene on SiO$_2$ or hBN, which typically has spin relaxation times of hundreds of ps to a few ns \cite{Tombros2007, Tombrosa2008, Jozsa2009, Han2011, Avsar2011, Zomer2012, Guimaraes2012, Neumann2013, Guimaraes2014, Drogeler2014, Kamalakar2015, Singh2016, Drogeler2016, Drogeler2017}. Comparable results were also observed in graphene/WS$_2$ heterostructures, where $\tau_\text{s}^\parallel$ was measured to be around 20 ps \cite{Omar2017}. However, the spin lifetime in the region of these devices without the TMDC was only around 40 ps. Thus, while these low spin lifetimes are suggestive of enhanced SOC in graphene, they are not experimental proof of this effect. As discussed in Section \ref{sec_theory_relaxation}, one of the consequences of the SOC induced in graphene by TMDCs is an unconventional large spin lifetime anisotropy \cite{Cummings2017}. Therefore, an experimental measurement of this phenomenon would provide strong evidence of proximity-induced SOC in graphene.

In two very recent experiments, a careful Hanle analysis was performed on graphene/MoSe$_2$ and graphene/WSe$_2$ at 75 K \cite{Ghiasi2017}, and on graphene/WS$_2$ and graphene/MoS$_2$ at room temperature \cite{Benitez2017}. The first set of measurements was performed using the in-plane magnetic field technique described in Section \ref{sec_hanle_inplane}, while the room temperature results were obtained using both this technique and oblique magnetic fields. In both experiments, the Hanle precession curves have shapes that are indicative of large spin lifetime anisotropy when the applied magnetic field allows the spins to rotate out of the graphene plane. Analysis of the measurements resulted in anisotropies of $\zeta \ge 10$ for all substrates and temperatures. This is shown in Fig.\ \ref{fig:hanle-experiments}, where the modeled curves for the in-plane and the oblique-field Hanle measurements are shown in panels (a) and (b), respectively, while the the corresponding measurements are in panels (c) and (d).

In these measurements, the graphene/WSe$_2$ heterostructure shows a giant anisotropy of $\zeta \approx 40$, in contrast to $\zeta \approx 11$ in graphene/MoSe$_2$. This difference in anisotropy is consistent with theory, which predicts a larger anisotropy for graphene on tungsten-based TMDCs compared to molybdenum-based TMDCs \cite{Gmitra2016, Cummings2017}. On the other hand, the measurements at room temperature of both graphene/WS$_2$ and graphene/MoS$_2$ exhibit $\zeta \approx 10$. One possible explanation for the lesser anisotropy of the WS$_2$ heterostructure could be the increased intravalley scattering due to phonons at room temperature, although it is not clear why this would impact WS$_2$ more than MoS$_2$. Another possibility is the quality of the graphene/TMDC interface. Although graphene/WS$_2$ has theoretically the highest $\lambda_\text{VZ} / \lambda_\text{R}$, the distance between the graphene and the TMDC can significantly impact the induced SOC strength \cite{Gmitra2016, Wang2015, Yang2016}, as discussed in Section \ref{sec_devices}. With a weaker graphene/TMDC interaction, other isotropic sources of spin relaxation, such as magnetic impurities or contact dephasing, could start to play a larger role and reduce the overall anisotropy. This could also explain some of the discrepancies between the measurements and the theoretical model \cite{Cummings2017}, which predicts values of $\zeta$ ranging from 20 to 200. Nevertheless, unprecedented spin lifetime anisotropy is clearly seen at room temperature, which is a crucial condition for implementing such heterostructures in spintronic devices. In addition, the oblique field measurements in sulphur-based TMDCs at different gate voltages also showed an on/off behavior in the spin signal \cite{Benitez2017}, in agreement with the aforementioned MoS$_2$-based devices. Meanwhile, this effect was not observed in selenium-based TMDCs, possibly due to their different band alignments \cite{Gmitra2016}.

Finally, the modulation of the spin lifetime has recently been measured in a double-gated graphene/WS$_2$ heterostructure \cite{Omar2017b}. The combination of a top and bottom gate allows one to maintain the Fermi level while tuning the electric field, which directly controls the Rashba SOC strength \cite{Gmitra2016}. By doing so, a fourfold modulation of $\tau_\text{s}^\parallel$ was observed as well as indications that $\zeta < 1$ in the strong Rashba regime. Moreover, the spin relaxation mechanism is predicted to be DP. Hence, these results not only support the microscopic theory of spin relaxation presented above, but also demonstrate an all-electric route to tune proximity effects in graphene. In conclusion, the Hanle experiments demonstrate that one can tune both the proximity effects and the spin transport in graphene by choosing the appropriate TMDC substrate. It is clear that strong SOC is induced in graphene no matter which TMDC is utilized, but also that the valley-Zeman SOC and intervalley scattering play an important role in these systems \cite{Cummings2017}.

%----------------------------------------FIGURE-----------------------------------------%
%---------------------------------------------------------------------------------------%

%%%%%%%%%%%%%%%%%%%SECTION%%%%%%%%%%%%%%%%%
\section{Weak Antilocalization in graphene/TMDC heterostructures} \label{sec_wal}
%%%%%%%%%%%%%%%%%%%%%%%%%%%%%%%%%%%%%%%%

In addition to directly measuring the spin signal, as described in Section \ref{spin-precession}, it is also possible to use charge transport measurements to glean information about the spin relaxation mechanisms in a material. This is done by measuring the weak localization and weak antilocalization phenomena, which are quantum corrections to the classical conductivity and are summarized in Fig.~\ref{fig_wal_theory}. In a diffusive system, charge scattering leads to a randomized path as an electron passes through a material. Some of these paths involve closed loops that result in backscattering, which can be traversed in either direction, as shown in Fig.~\ref{fig_wal_theory}(a). Owing to the wave nature of electrons, these two opposite loops can constructively interfere with one another, increasing the weight of this backscattering path and reducing the overall conductivity compared to the classical case \cite{Abrahams1979}. This is weak localization, and is illustrated in Fig.~\ref{fig_wal_theory}(b). In the presence of spin-orbit coupling, the phase between these opposite loops switches sign, leading to destructive interference. This reduces the weight of the backscattering loop and increases the conductivity relative to the classical value \cite{Hikami1980}. This is the process known as weak antilocalization, and is shown in Fig.~\ref{fig_wal_theory}(c). By applying a magnetic field, one can tune the phase accumulated along each diffusive path, and thus alter the interference properties of the backscattering loops. This will modulate the conductivity, and its precise dependence on magnetic field can be used to infer the strength of the various charge and spin relaxation mechanisms in the measured system \cite{Altshuler1980, Hikami1980}.

%%%%%%%%%%%%%%%%%%%SUBSECTION%%%%%%%%%%%%%%%%%
\subsection{WAL Theory} \label{sec_wal_theory}
%%%%%%%%%%%%%%%%%%%%%%%%%%%%%%%%%%%%%%%%
A comprehensive description of WAL in graphene was developed by McCann and Fal'ko \cite{Mccann2012}, who derived the quantum correction to the conductivity as
\begin{equation} \label{eq_walfull}
\begin{split}
\Delta \sigma(B) &\equiv \sigma(B) - \sigma(0) \\
&= - \frac{e^2}{2\pi h} \sum \limits_{j,l = 0,x,y,z} {c_j c_l F \left( \frac{\tau_B^{-1}}{\tau_\phi^{-1} + \Gamma_j^l} \right)},
\end{split}
\end{equation}
where $F(z) = \ln(z) + \Psi(1/2 + 1/z)$, $\Psi$ is the digamma function, $c_0 = 1$, $c_x = c_y = c_z = -1$,  $\tau_{B}^{-1} = 4DeB/\hbar$, $D$ is the diffusion coefficient, $B$ is the external magnetic field perpendicular to the graphene plane, and $\tau_\phi$ is the inelastic dephasing time. The $\Gamma_j^l$ contain the various relaxation rates in the system that arise from symmetry breaking, including both spin and charge relaxation, and are listed in Table \ref{table_wal}. The subscripts of $\Gamma_j^l$ refer to spin relaxation and the superscripts refer to valley relaxation. The rates $\tau_\text{I}^{-1}$, $\tau_\text{VZ}^{-1}$, $\tau_\text{R}^{-1}$, and $\tau_\text{PIA}^{-1}$ are, respectively, the spin relaxation rates arising from intrinsic, valley-Zeeman, Rashba, and PIA SOC. The rate $\tau_*^{-1} = \tau_\text{iv}^{-1} + \tau_\text{z}^{-1}$, where $\tau_\text{z}$ is an intravalley scattering time related to fluctuations in the onsite energy of the graphene A/B sublattices and fluctuations in the nearest-neighbor hopping. For interference effects to occur, the electrons must maintain phase coherence around the scattering loops; in general the relation $\tau_\phi^{-1} < \Gamma_j^l$ satisfies this condition.

\begin{table}[t]
\centering
\begin{tabular}{c c c}
\toprule
\multicolumn{3}{c}{Relaxation rates} \\
\midrule
$\Gamma_0^0$ & = & $0$ \\
$\Gamma_0^x = \Gamma_0^y$ & = & $\tau_*^{-1} + \tau_\text{VZ}^{-1}$ \\
$\Gamma_0^z$ & = & $2\tau_\text{iv}^{-1}$ \\
$\Gamma_x^0 = \Gamma_y^0$ & = & $\tau_\text{R}^{-1} + \tau_\text{PIA}^{-1} + \tau_\text{I}^{-1} + \tau_\text{VZ}^{-1}$ \\
$\Gamma_x^x = \Gamma_x^y = \Gamma_y^x = \Gamma_y^y$ & = & $\tau_*^{-1} + \tau_\text{R}^{-1} + \tau_\text{PIA}^{-1} + \tau_\text{I}^{-1}$ \\
$\Gamma_x^z = \Gamma_y^z$ & = & $2\tau_\text{iv}^{-1} + \tau_\text{R}^{-1} + \tau_\text{PIA}^{-1} + \tau_\text{I}^{-1} + \tau_\text{VZ}^{-1}$ \\
$\Gamma_z^0$ & = & $2\tau_\text{R}^{-1}$ \\
$\Gamma_z^x = \Gamma_z^y$ & = & $\tau_*^{-1} + 2\tau_\text{R}^{-1} + 2\tau_\text{PIA}^{-1} + \tau_\text{VZ}^{-1}$ \\
$\Gamma_z^z$ & = & $2\tau_\text{iv}^{-1} + 2\tau_\text{R}^{-1} + 2\tau_\text{PIA}^{-1}$ \\
\bottomrule
\end{tabular}
\caption{Relaxation rates contributing to the quantum correction to the conductivity, Eq.\ (\ref{eq_walfull}), here we only consider uniform SOC and neglect the contribution of spin-orbit disorder. Adapted from ref.\ \citenum{Mccann2012} with permission from the American Physical Society, \textcopyright 2012. and updated to include PIA and valley-Zeeman SOC}
\label{table_wal}
\end{table}

Eq.~(\ref{eq_walfull}) contains a total of 16 terms, but is typically simplified by assuming that the intervalley scattering rate, $\tau_\text{iv}^{-1}$, is much faster than the spin relaxation rates. Because $F(z)$ is a monotonically increasing function of $z$, all terms containing $\tau_\text{iv}^{-1}$ can be neglected, leaving
\begin{align} \label{eq_wal}
\Delta \sigma(B) = &- \frac{e^2}{2\pi h} \left[ F \left( \frac{\tau_B^{-1}}{\tau_{\phi}^{-1}} \right)
- F \left( \frac{\tau_{B}^{-1}}{\tau_{\phi}^{-1}+2\tau_\text{asy}^{-1}} \right) \right. \nonumber \\
&- 2F \left. \left( \frac{\tau_{B}^{-1}}{\tau_{\phi}^{-1}+\tau_\text{asy}^{-1}+\tau_\text{sym}^{-1}} \right) \right].
\end{align}
Here, $\tau_\text{asy}$ is the spin relaxation time arising from symmetry breaking perpendicular to the graphene plane. This is typically associated with Rashba SOC but should also include PIA SOC, such that $\tau_\text{asy}^{-1} = \tau_\text{R}^{-1} + \tau_\text{PIA}^{-1}$. Meanwhile, $\tau_\text{sym}$ is the spin relaxation time associated with SOC that maintains perpendicular symmetry, which is usually ascribed to intrinsic, or Kane-Mele \cite{KaneMele2005}, SOC. However, in graphene/TMDC systems the valley-Zeeman SOC should also be considered, giving $\tau_\text{sym}^{-1} = \tau_\text{I}^{-1} + \tau_\text{VZ}^{-1}$. The total spin relaxation time is then defined as $\tau_\text{so}^{-1} \equiv \tau_\text{asy}^{-1} + \tau_\text{sym}^{-1}$.

\begin{figure}[t]
\includegraphics[width=\columnwidth]{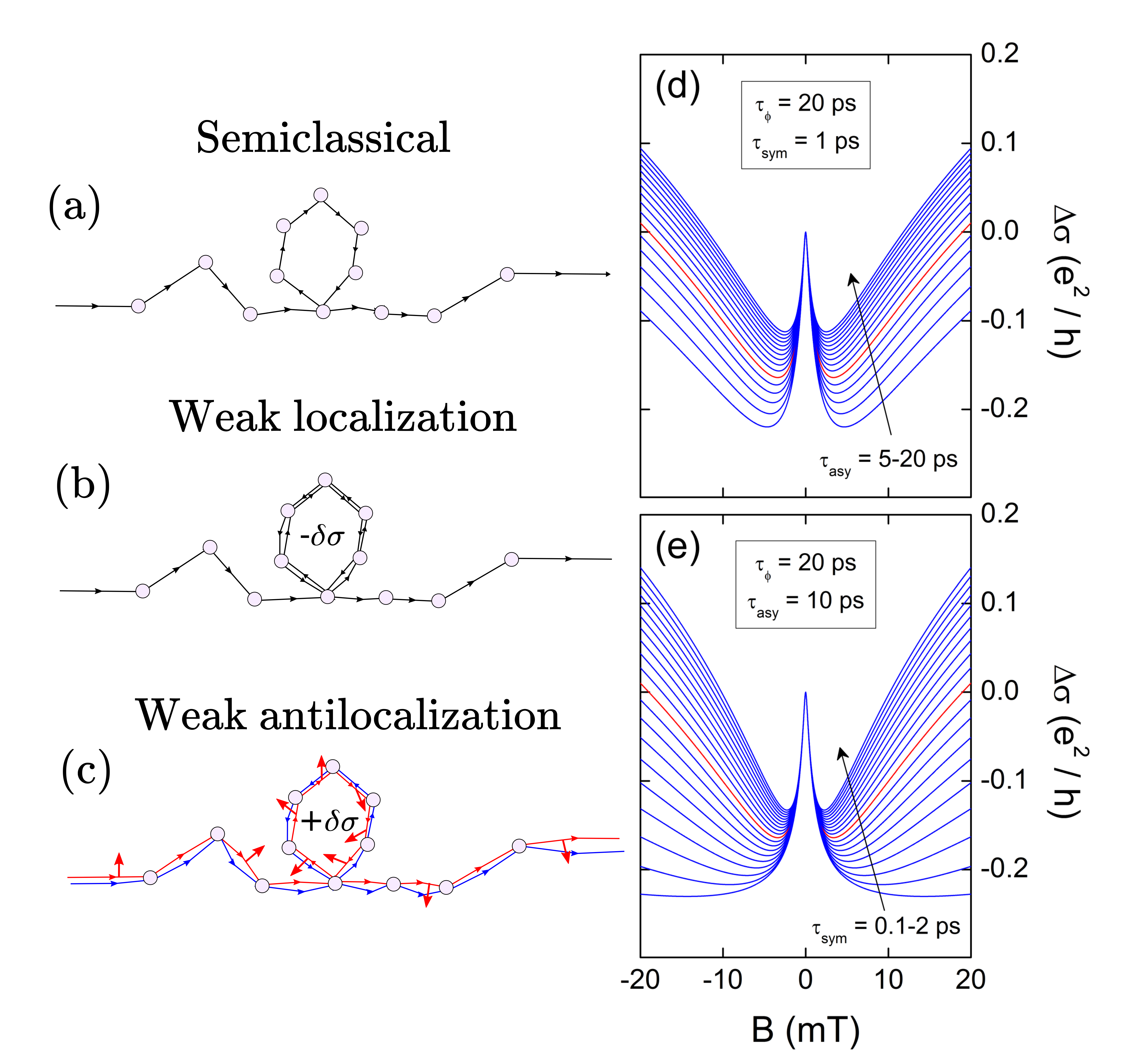}
\caption{Summary of WAL in diffusive materials. Panel (a) depicts charge transport in the semiclassical regime, including a pair of backscattering loops. Panels (b) and (c) illustrate the case in the WL and WAL regimes, where the backscattering loops constructively or destructively interfere, respectively. Panels (d) and (e) show magnetoconductance curves using typical relaxation times for graphene/TMDC heterostructures.}
\label{fig_wal_theory}
\end{figure}

Figures \ref{fig_wal_theory}(d) and (e) show examples of the magnetoconductivity and how it depends on $\tau_\text{asy}$ and $\tau_\text{sym}$. The red curves show the case for a baseline set of parameters that are typical of those measured in graphene/TMDC heterostructures: $\tau_\phi = 20$ ps, $\tau_\text{asy} = 10$ ps, and $\tau_\text{sym} = 1$ ps. In Fig.~\ref{fig_wal_theory}(d) we vary $\tau_\text{asy}$ between 5 and 20 ps, and in this range one can see that the primary effect is to scale the height of the central WAL peak, with smaller $\tau_\text{asy}$ yielding a larger peak. In Fig.~\ref{fig_wal_theory}(e) we vary $\tau_\text{sym}$ between 0.1 and 2 ps, and find that its primary role is to adjust the slope of $\Delta \sigma$ at higher magnetic fields, with a flat profile indicative of a small $\tau_\text{sym}$.
%----------------------------------------FIGURE-----------------------------------------%
%-----------------------------------------------------------------------------------------%

%%%%%%%%%%%%%%%%%%%SUBSECTION%%%%%%%%%%%%%%%%%
\subsection{Measurements of WAL in graphene/TMDC heterostructures} \label{sec_wal_measurements}
%%%%%%%%%%%%%%%%%%%%%%%%%%%%%%%%%%%%%%%%%%
The first measurements of WAL in a graphene/TMDC heterostructure were made in graphene on a WS$_2$ substrate \cite{Wang2015}. After averaging out conductance fluctuations, the magnetoconductance curves exhibited large WAL peaks around $B=0$, indicating strong SOC induced in graphene by the WS$_2$ substrate. Fits to Eq.\ (\ref{eq_wal}) yielded $\tau_\text{so} = 2.5\text{--}5$ ps, and $\tau_\text{asy}$ was estimated to be approximately 3$\times$ larger. Although the role of valley-Zeeman SOC was not considered, the authors noted a strong correlation between $\tau_\text{so}$ and $\tau_\text{iv}$. This was attributed to spatial fluctuations of the spin-orbit coupling strength, which were posited to simultaneously cause intervalley scattering and relax the spin.

The next reported measurement also considered a graphene/WS$_2$ device \cite{Yang2016}. In the WAL analysis, the impact of intrinsic SOC on $\tau_\text{sym}$ was assumed to be negligible, as was the valley-Zeeman SOC. This left $\tau_\phi$ and $\tau_\text{asy}$ as the only fitting parameters in Eq.\ (\ref{eq_wal}), from which $\tau_\text{asy} \approx 5$ ps was extracted. This relaxation time was found to scale inversely with the momentum scattering time, and could also be tuned by $\sim$10\% in either direction with a vertical electrical field. Both of these observations are suggestive of the DP mechanism of spin relaxation induced by Rashba SOC, with an estimated Rashba strength of $\lambda_\text{R} = 0.4$ meV. A follow-up work by the same group \cite{Yang2017} found similar results for graphene interfaced with WSe$_2$ and MoS$_2$. Using the same analysis they found $\tau_\text{asy} = 2\text{--}10$ ps, which scaled inversely with $\tau_\text{p}$, giving $\lambda_\text{R} = 1.5 \, (0.9)$ meV for graphene interfaced with WSe$_2$ (MoS$_2$).

An independent set of measurements of graphene/WSe$_2$ also concluded that spin relaxation was dominated by the traditional DP mechanism, but considered both $\tau_\text{asy}$ and $\tau_\text{so}$ in the WAL fits \cite{Volkl2017}. These fits yielded $\tau_\text{asy} = 1.7 \text{--} 4.5$ ps and $\tau_\text{so} = 0.9 \text{--} 1.5$ ps, and by examining the relationship between $\tau_\text{so}$ and $\tau_\text{p}$, a Rashba spin-orbit strength of $\lambda_\text{R} = 0.7 \text{--} 1$ meV was extracted.

\begin{figure}[t]
\includegraphics[width=\columnwidth]{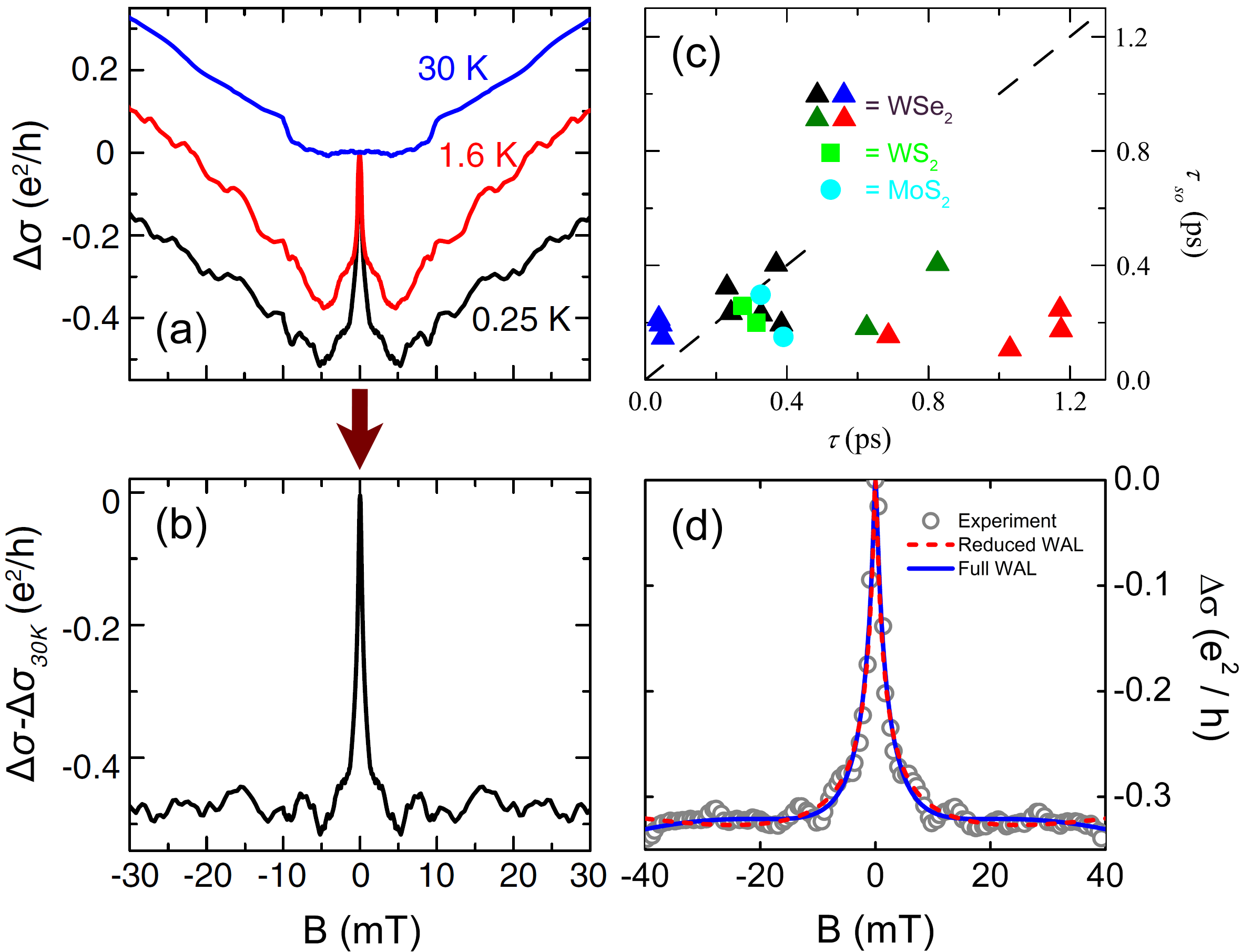}
\caption{(a) Temperature-dependent evolution of the magnetoconductance in a graphene/WS$_2$ heterostructure. Subtracting the high-temperature classical signal gives the normalized magnetoconductivity shown in (b). Panel (c) shows a summary of WAL measurements made using this background subtraction, yielding a uniform spin-orbit time across a wide range of devices (symbol colors) and TMDCs (symbol shapes). Panels (a)-(c) are reproduced from ref.\ \citenum{Wang2016} . with permission from the American Physical Society, \textcopyright 2016. (d) Fits to magnetoconductance using the full and the reduced WAL formulas of Eqs. (\ref{eq_walfull}) and (\ref{eq_wal}), with experimental data extracted from ref.\ \citenum{Wakamura2017}. with permission from the American Physical Society, \textcopyright 2018}
\label{fig_wal_exp}
\end{figure}

Meanwhile, a comprehensive set of WAL measurements demonstrated the importance of eliminating classical effects from the magnetoconductance \cite{Wang2016}. At high temperatures the dephasing time $\tau_\phi$ becomes very short, washing out interference effects, and any dependence of the conductivity on the magnetic field can be considered to arise from classical effects. In these measurements, subtracting the high-temperature magnetoconductivity curves from the low-temperature curves resulted in a sharp WAL peak and little to no upturn of $\Delta \sigma$ at higher fields; see Figs.\ \ref{fig_wal_exp}(a) and (b) for an example. As shown in Fig.\ \ref{fig_wal_theory}(e), this flat profile of $\Delta \sigma$ is indicative of very fast spin relaxation. Indeed, fits to WAL theory yielded upper bounds of $\tau_\text{so} \le 0.1\text{--}0.4$ ps. As shown in Fig.\ \ref{fig_wal_exp}(c), this behavior was found over many devices, including different TMDCs (MoS$_2$, WS$_2$, and WSe$_2$) and a wide range of mobilities ($3000 \text{--} 110\,000$ cm$^2$/V$\cdot$s). Although small values of $\tau_\text{so}$ were reported, the individual values of $\tau_\text{asy}$ and $\tau_\text{sym}$ were not. Finally, through an analysis of Shubnikov-de Haas oscillations in bilayer graphene/WSe$_2$ devices, the Rashba SOC strength was estimated to be $\lambda_\text{R} = 10\text{--}15$ meV, which is one order of magnitude larger than what was found in other measurements or in {\it ab initio} simulations \cite{Gmitra2016, Wang2016, Yang2016}.

While all of these measurements demonstrate significant proximity-induced SOC in graphene, the analyses do not consider the impact of valley-Zeeman SOC. As predicted theoretically \cite{Cummings2017}, and supported by Hanle measurements \cite{Ghiasi2017, Benitez2017}, valley-Zeeman SOC is responsible for fast relaxation of the in-plane spins and a resultant giant spin lifetime anisotropy. As shown very recently, a careful analysis of WAL measurements can reveal this behavior \cite{Zihlmann2018}. According to the theory of McCann and Fal'ko, $\tau_\text{sym}$ is determined by both intrinsic and valley-Zeeman SOC, while $\tau_\text{asy}$ is dominated by Rashba SOC \cite{Mccann2012}. As discussed in Refs.\ \citenum{Yang2016, Yang2017, Volkl2017, Zihlmann2018}, the contribution of intrinsic SOC is negligible, leaving valley-Zeeman SOC as the only contributor to $\tau_\text{sym}$. By connecting Eq.\ (\ref{eq:rates_intra}) to Table \ref{table_wal}, one can then assign $\tau_\text{s}^\perp = \tau_\text{asy} / 2$ and $\tau_\text{s}^\parallel = \tau_\text{so}$, which allows a determination of both $\lambda_\text{R}$ and $\lambda_\text{VZ}$. Additionally, the spin lifetime anisotropy can be estimated as $\zeta = \tau_\text{asy} / 2\tau_\text{sym}$. In recent measurements of graphene/WSe$_2$ devices \cite{Zihlmann2018}, this analysis resulted in $\lambda_\text{R} = 0.35$ meV, $\lambda_\text{VZ} = 0.2\text{--}2$ meV, and $\zeta = 20$. These results match well with DFT calculations \cite{Wang2015, Yang2016, Gmitra2016}, spin relaxation theory \cite{Cummings2017}, and the Hanle measurements of spin lifetime anisotropy \cite{Ghiasi2017, Benitez2017} and thus highlight the importance of considering the valley-Zeeman SOC when studying quantum transport in these systems. As mentioned previously, to date there has been no experimental confirmation of the presence of PIA SOC, and thus it has been neglected in the WAL analysis.

%----------------------------------------FIGURE-----------------------------------------%
%----------------------------------------------------------------------------------------%

%%%%%%%%%%%%%%%%%%%SUBSECTION%%%%%%%%%%%%%%%%%
\subsection{WAL analysis in the strong-SOC regime} \label{sec_wal_strong_soc}
%%%%%%%%%%%%%%%%%%%%%%%%%%%%%%%%%%%%%%%%%%

As mentioned in Section \ref{sec_wal_theory}, Eq.\ (\ref{eq_wal}) was derived assuming the strong-scattering/weak-SOC regime, i.e., in the limit $\tau_\text{iv} \ll \tau_\text{so}$. However, owing to graphene's exceptional charge transport properties and the strong SOC induced by the TMDC, this condition can be violated in graphene/TMDC heterostructures. Indeed, this appears to be the case in at least one of the aforementioned WAL measurements \cite{Wang2016}, where $\tau_\text{so} \approx \tau_\text{p}$. This then raises the question, are the fits obtained from Eq.\ (\ref{eq_wal}) reasonable, or is a more general fit using Eq.\ (\ref{eq_walfull}) required?

The full WAL equation requires a fit of six independent time scales: $\tau_\phi$, $\tau_\text{iv}$, $\tau_\text{z}$, $\tau_\text{asy}$, $\tau_\text{I}$, and $\tau_\text{VZ}$. Finding a unique fit is much more difficult in this regime, and a measurement that captures the tails of the magnetoconductance at higher magnetic fields is crucial. Therefore, we consider a very recent measurement of magnetoconductivity out to $\pm 40$ mT in a graphene/WS$_2$ system \cite{Wakamura2017}, displayed as the open circles in Fig.\ \ref{fig_wal_exp}(d). The mobility of this sample is $\mu = 12\,000$ cm$^2$/V$\cdot$s, giving $\tau_\text{p} \approx 0.1$ ps. The dashed red line in Fig.\ \ref{fig_wal_exp}(d) shows the fit using the reduced WAL theory of Eq.\ (\ref{eq_wal}), which yields $\tau_\text{asy} = 3.6$ ps and $\tau_\text{sym} = 0.05$ ps. Because $\tau_\text{sym} < \tau_\text{p}$, Eq.\ (\ref{eq_wal}) is clearly beyond its range of validity. The solid blue line shows the fit to the full WAL formula. In this fit, there remains a fair amount of flexibility in the values of $\tau_\text{iv}$ and $\tau_\text{z}$, all of which give nearly identical magnetoconductance curves. By varying $\tau_\text{iv}$ and $\tau_\text{z}$ in the ranges $\tau_\text{iv} \in [2,15]\tau_\text{p}$ and $\tau_\text{z} \in [0.5,2]\tau_\text{p}$, we find $\tau_\text{asy} = 1.6\text{--}2.5$ ps and $\tau_\text{sym} = 0.3\text{--}0.5$ ps. This analysis shows that when the conditions of Eq.\ (\ref{eq_wal}) are violated, considering the full WAL theory can have a considerable impact on the estimate of $\tau_\text{sym}$, increasing it by a factor of 6 to 10 in this particular case. Looking forward, an independent measurement of $\tau_\text{iv}$ and $\tau_\phi$ would make for an easier fit to Eq.\ (\ref{eq_walfull}). We should highlight however that, although in Fig. 7 we show data of Ref. \cite{Wakamura2017} that violates the standard assumptions of WAL theory, the authors have other samples were those conditions were satisfied and similar scaling behaviors were obtained, thus the difficulty seem not to be crucial for the determination of the relaxation regime.

%%%%%%%%%%%%%%%%%%%SECTION%%%%%%%%%%%%%%%%%
\section{Spin Hall Effect} \label{sec_she}
%%%%%%%%%%%%%%%%%%%%%%%%%%%%%%%%%%%%%%%

Up to this point, we have focused solely on the impact that TMDCs have on spin relaxation in graphene. However, for low-power spintronics applications it is also necessary to have a means of generating spin-polarized currents. The spin Hall effect is one way of doing this, as it is a phenomenon by which an electric current produces a transverse spin current as a result of spin-orbit coupling \cite{RevModPhys.87.1213}. Depending on its origin, the spin Hall effect is classified as either extrinsic or intrinsic. The extrinsic spin Hall effect is driven by charge scattering in the presence of SOC, which allows for spin-up and spin-down electrons to be scattered in opposite directions. The intrinsic spin Hall effect, on the other hand, is generated by the homogeneous SOC field present in the material, which acts as a spin-dependent Lorentz force that pushes spin-up and spin-down electrons in opposite directions. In Section \ref{sec_theory_model} we showed that TMDCs induce nontrivial SOC in graphene, with valley-Zeeman being the dominant term. We will show in this section that this SOC field may also generate a relatively large spin Hall effect \cite{Garcia2017}. Moreover, guided by the theoretical framework of the previous sections, we will discuss the conditions needed to measure the intrinsic SHE in graphene/TMDC systems. Finally, it is important to note that the intrinsic mechanism may not be the only source of spin Hall effect, as chalcogenide vacancies could produce the extrinsic SHE \cite{Avsar2014}. However, given that up to now there is no clear evidence of this effect, we will focus on the intrinsic SHE.

%%%%%%%%%%%%%%%%SUBSECTION%%%%%%%%%%%%%%%%%
\subsection{Topology and spin Hall Effect in graphene/TMDCs}
%%%%%%%%%%%%%%%%%%%%%%%%%%%%%%%%%%%%%%%

In previous sections, we have shown that treating the SOC as an effective magnetic field is sufficient to quantitatively explain the various experimental and numerical results. Similarly, one may argue that this effective magnetic field should also affect the orbital motion of electrons, giving rise to a spin-dependent Hall effect. This is indeed the case, although the connection between the spin Hall effect and the spin-orbit coupling is not straightforward.

To visualize this, consider a system in equilibrium with no net spin or charge current. Under such a situation the spins are aligned with the effective magnetic field, as illustrated in Fig.\ \ref{fig:rashba-vs-vz}. After the application of an external electric field $\bm{E}$, the system undergoes two changes: first, the electrons acquire a finite momentum along the direction of $\bm{E}$, and second, the wave functions $\phi_n(\bm{k})$ of the system are modified, and with them the expectation values of all observables. For example, the velocity operator takes the form

\begin{equation}
\bm{v}_n(\bm{k})= \bm{\nabla}\varepsilon_n(\bm{k})-\frac{e}{\hbar}\bm{E}\times 
\bm{\Omega}_n(\bm{k}), \label{anomalous-vel}
\end{equation} 
where $n$ is the band index, $\bm{k}$ is the electron momentum, $\varepsilon_n(\bm{k})$ is the energy of band $n$, and
$\bm{\Omega}_n(\bm{k}) = \bm{\nabla_k} \times \langle \phi_n(\bm{k}) | \nabla_{\bm{k}} | \phi_n(\bm{k}) \rangle$ is known as the Berry curvature, a quantity closely related to the effect of the spin-orbit coupling on the topological properties of the electron's spectrum \cite{RevModPhys.87.1213, Xiao2010}. The first term in Eq.\ (\ref{anomalous-vel}) describes the movement of electrons along the direction of $\bm{E}$. The second term shows that under the action of an electrical field, the velocity acquires a perpendicular contribution directly related to the wave functions of the underlying material. This term is the origin of the different Hall effects that may arise in a given system \cite{Thouless1982, CrestiRNC2016}. When it originates from certain types of spin-orbit interactions, due to time-reversal symmetry it will have opposite sign for opposite spins, giving rise to a transverse spin current, i.e., the spin Hall effect.

In spintronic devices, the figure of merit for the spin Hall effect is the spin Hall angle 
\begin{equation}
\gamma_{\text{sH}}\equiv \frac{\sigma_{xy}^z}{\sigma_{xx}},
\end{equation}
where $\sigma_{xx}$ is the longitudinal charge conductivity assuming an electric field along the $x$-direction, $\bm{E} = E_x \bm{x}$, and $\sigma_{xy}^z$ is the spin Hall conductivity defined through the relation $j_{s,\beta}^{z} =\sigma_{\alpha \beta}^z E_\alpha$, where $j_{s,\beta}^{z}$ is the $\beta$ component of the spin current polarized in the out-of-plane direction \cite{CrestiRNC2016}. The spin current can be determined microscopically as the expectation value of the spin current operator $\hat{\jmath}_\beta^{z}$, which for a system out of equilibrium is computed as \cite{RevModPhys.87.1213, Garcia2016, CrestiRNC2016}
\begin{equation}
j_{s,\beta}^{z}=\langle\hat{\jmath}_\beta^{z}\rangle\equiv\text{Tr}\left[\hat{\jmath}_\beta^{z} \hat{\rho}_{\text{neq}}(\bm{E}) \right], \label{Kubo-Basin}
\end{equation}
where $\hat{\rho}_{\text{neq}}(\bm{E})$ the density matrix driven out of equilibrium by the applied electric field $\bm{E}$ and 
\begin{equation}
\hat{\jmath}_\beta^{z}\equiv -\frac{e}{2\hbar V}\{s_z,\hat{v}_\beta\}
\end{equation}
is the spin current operator, with the curly brackets representing the anticommutator between the $z$-component of the spin and the $\beta$-component of the velocity operator, and $V$ the volume of the sample. The nonequilibrium density matrix can be computed exactly for noninteracting electrons using a variety of techniques such as Landauer-B\"{u}ttiker \cite{Datta1995} or the Kubo formalism\cite{Kubo1957,Mahan2000}, or can be approximated by solving the Boltzmann equation \cite{Datta1995}. Throughout this section, our results were obtained via an efficient numerical evaluation of the Kubo formula, which allows for the simulation of realistic system sizes and yields results compatible with other predictions \cite{Garcia2015, Garcia2016, Garcia2017}.

\begin{figure}[h!]
\centering
\includegraphics{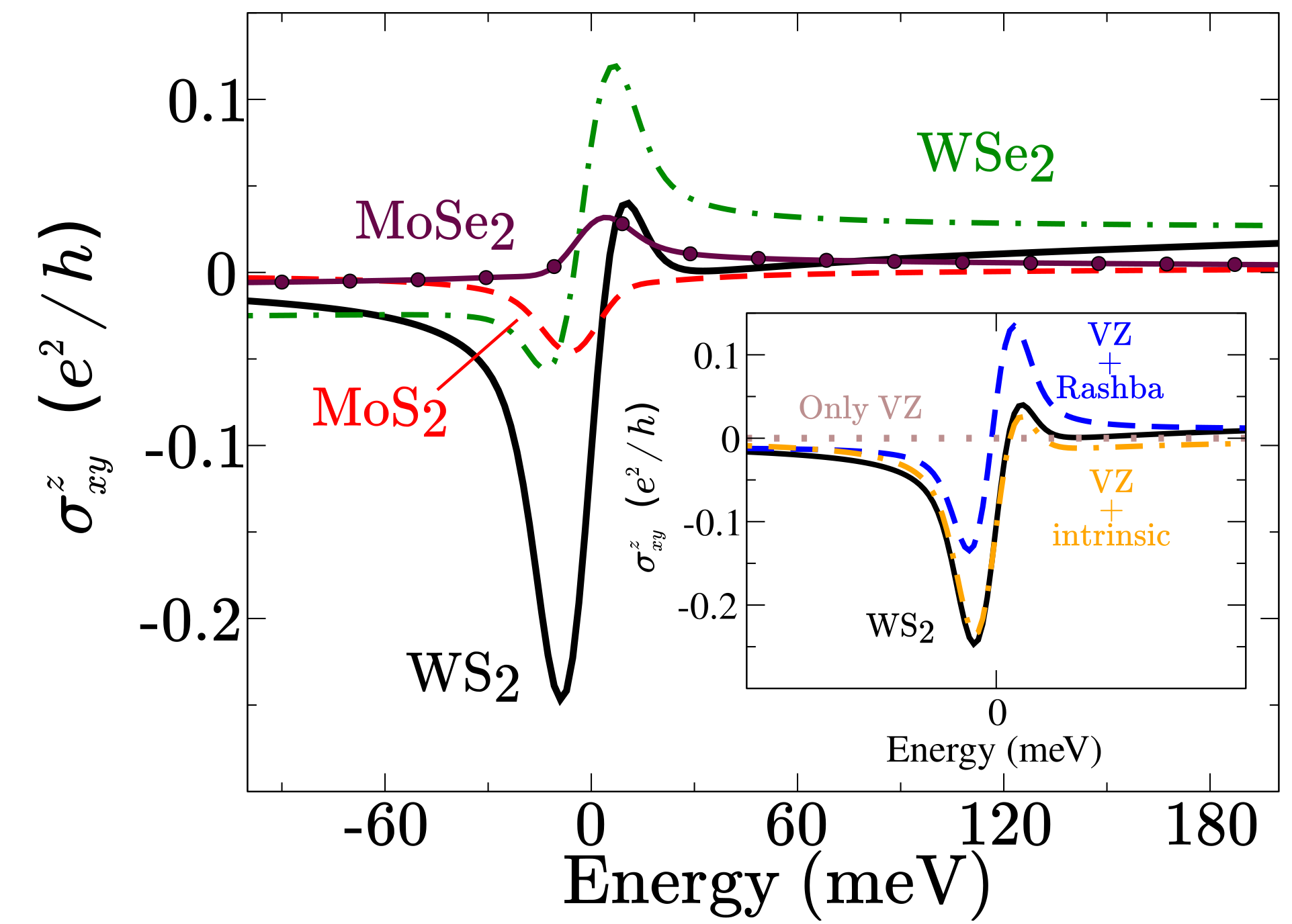}
\caption{Spin Hall conductivity for pristine graphene on different TMDCs, with graphene/TMDC model parameters shown in Table \ref{table_params}. In the inset we show the effect of different SOC combinations on the spin Hall conductivity of WS$_2$. Data adapted from ref.\ \citenum{Garcia2017} with permission from the American Chemical Society \textcopyright 2017. }
\label{fig:pristineSpinHall}
\end{figure}

Given that the spin Hall conductivity is proportional to the spin Hall current, one would expect $\sigma_{{xy}}^{z}$ to be related to the Berry curvature. This is indeed the case; as demonstrated by Thouless et al. for the quantum Hall effect \cite{Thouless1982}, and later extended to spin for pristine systems, showing that  Eq.\ (\ref{Kubo-Basin}) is proportional to the Berry curvature \cite{RevModPhys.87.1213}. In this sense, the pristine spin Hall conductivity is a measure of a material's potential for generating transverse spin currents via the intrinsic spin Hall effect \cite{Dyrda2017}. In Fig.\ \ref{fig:pristineSpinHall} we show this quantity for different graphene/TMDC heterostructures as computed in Ref.\ \citenum{Garcia2017}. This calculation shows that all TMDCs can generate the intrinsic spin Hall effect, with WS$_2$ standing out as the most promising one for realizing this phenomenon. As shown in the inset, valley-Zeeman SOC alone does not lead to a spin Hall effect, and only in combination with Rashba and/or intrinsic SOC is the spin Hall conductivity finite. Finally, it should be stressed that disorder may have a major impact on the spin Hall conductivity when the system is in the diffusive regime, as we show in the next section.

%----------------------------------------FIGURE-----------------------------------------%
%-----------------------------------------------------------------------------------------%

%%%%%%%%%%%%%%%%SUBSECTION%%%%%%%%%%%%%%%%%
\subsection{The role of disorder}
%%%%%%%%%%%%%%%%%%%%%%%%%%%%%%%%%%%%%%%
A nonzero Berry curvature is a necessary but not sufficient condition for the existence of a proximity-induced SHE. This is because,
despite the presence of a finite intrinsic SHE in the clean limit, in the diffusive regime disorder can completely suppress the intrinsic
contribution. This phenomenon was first predicted in the context of Rashba 2DEGs, but it was later proven to be the case for
any k-linear SOC[54,143]. For Dirac systems in the diffusive regime an equivalent theorem was absent until the recent work of Milletari
et al., which arrived at the same conclusion: in graphene with Rashba SOC the intrinsic SHE is completely suppressed by
disorder [52]. However, in the same work it was shown that when valley-Zeeman SOC is included, the spin Hall current generated
by an electric field in the x direction will take the form
\begin{equation}
j_{s,y}^{z}=v_\text{F} \frac{\lambda_\text{VZ}}{\lambda_\text{R}}( n_\text{s}^{x,\text{K}}- n_\text{s}^{x,\text{K}'}), \label{MilEq}
\end{equation}
thus implying that as long as the macroscopic nonequilibrium spin density is slightly different at each valley, the spin Hall current will be finite.

\begin{figure}[h!]
\centering
\includegraphics[width=\columnwidth]{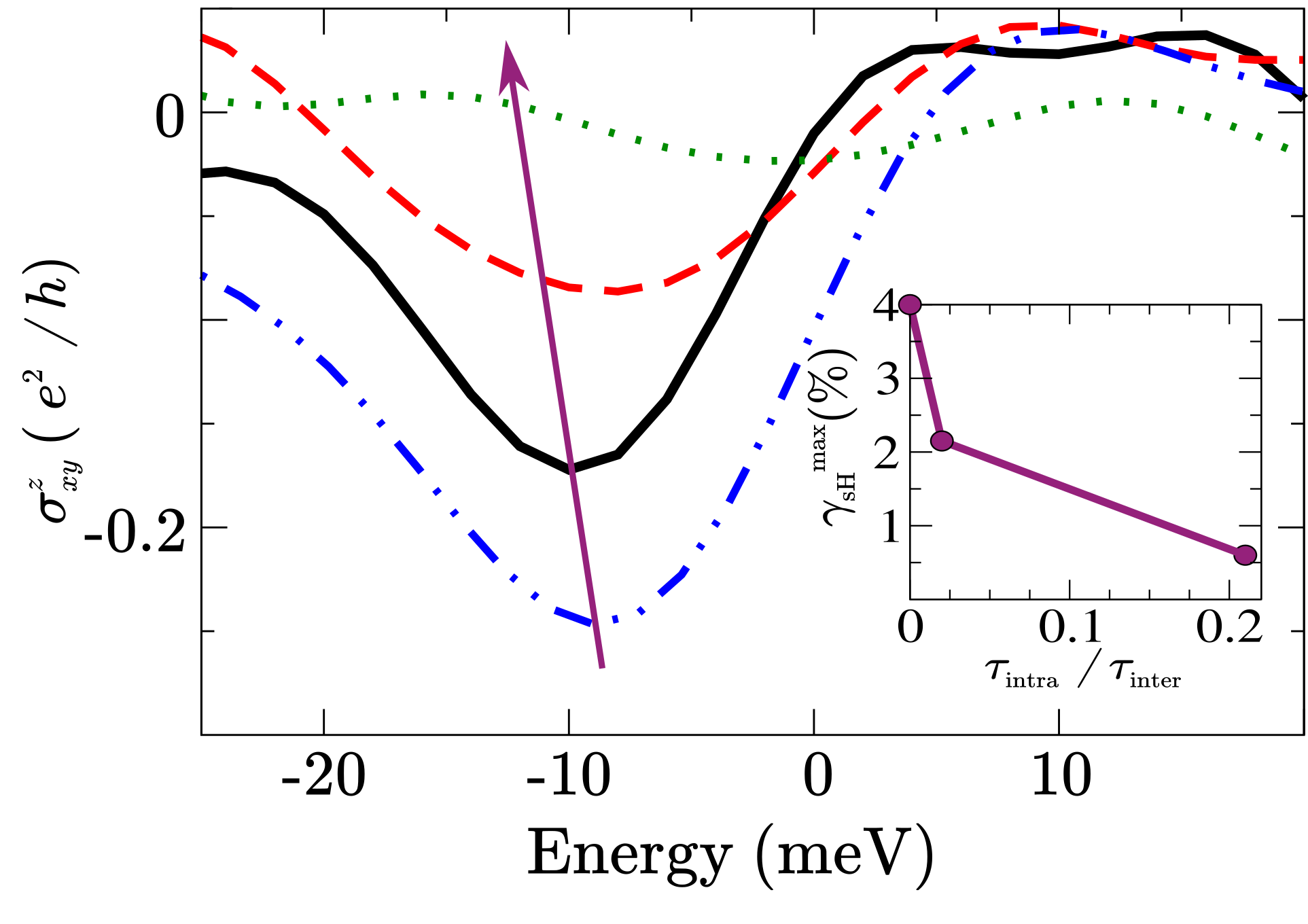}
\caption{Spin Hall conductivity of graphene/WS$_2$ for weak (solid black line), intermediate (dashed red line) and strong (dotted green line) intervalley scattering, with the pristine spin Hall conductivity (dot-dashed blue line) shown as a reference. The inset shows the scaling of the spin Hall angle with intervalley scattering strength. Graphene/TMDC model parameters shown in Table \ref{table_params}. Data adapted from ref.\ \citenum{Garcia2017} with permission from
American Chemical Society, \textcopyright 2017.	}
\label{fig:disorderSpinHall}
\end{figure}

The previous formula exemplifies once again that valley-Zeeman SOC plays an important role for the spin Hall effect, even in the presence of disorder.  However, Eq.\ (\ref{MilEq}) does not take into account intervalley scattering, which as we showed in previous sections crucially determines the strength of different phenomena. In Fig.\ \ref{fig:disorderSpinHall}, we show the spin Hall conductivity and spin Hall angle computed using Eq.\ (\ref{Kubo-Basin}) for disorder profiles with similar mobilities but very different intervalley scattering rates \cite{Garcia2017}. One can see that for weak intervalley scattering a large spin Hall angle is obtained, similar to that seen in pristine graphene and in agreement with Eq.\ (\ref{MilEq}). However, for stronger intervalley scattering the spin Hall conductivity is strongly suppressed, even for samples with similar mobilities and conductivities. A possible explanation for this relies on time-reversal symmetry, which implies a change of sign of the Berry curvature for each valley and would lead to zero average Berry curvature when intervalley scattering is strong enough \cite{Garcia2017}. This information is key for the design of experiments aiming to observe the SHE in graphene/TMDC systems, given that vacancies, grain boundaries, and other short-range defects will cause intervalley scattering and prevent the observation of the SHE. Therefore, as discussed in Section \ref{sec_devices}, there is a significant need for experimental techniques that can correlate the different sources of disorder with intervalley scattering, if one hopes to completely control the spin in these heterostructures.

%----------------------------------------FIGURE------------------------------------ltena-----%
%------------------------------------------------------------------------------------------%

%%%%%%%%%%%%%%%%SUBSECTION%%%%%%%%%%%%%%%%%
\subsection{Experimental measurement of the SHE}
%%%%%%%%%%%%%%%%%%%%%%%%%%%%%%%%%%%%%%%
Over the past several years, evidence of the spin Hall effect has been reported in graphene decorated with various adatoms \cite{Balakrishnan2013, Balakrishnan2014, Avsar2015}, and in graphene/TMDC heterostructures \cite{Avsar2014}. All these measurements utilized a purely electrical experiment proposed by Abanin et al \cite{Abanin2009}. However, this proposal was developed for measuring a nonlocal signal originating from the extrinsic spin Hall effect, and it is not clear that the same approach can be used for the intrinsic SHE. To measure the intrinsic SHE, one needs to create and transport spin currents in a material with large SOC, but in such materials coherent spin precession can dephase and suppress the signal at the detector.

\begin{figure}[h!]
\centering
\includegraphics{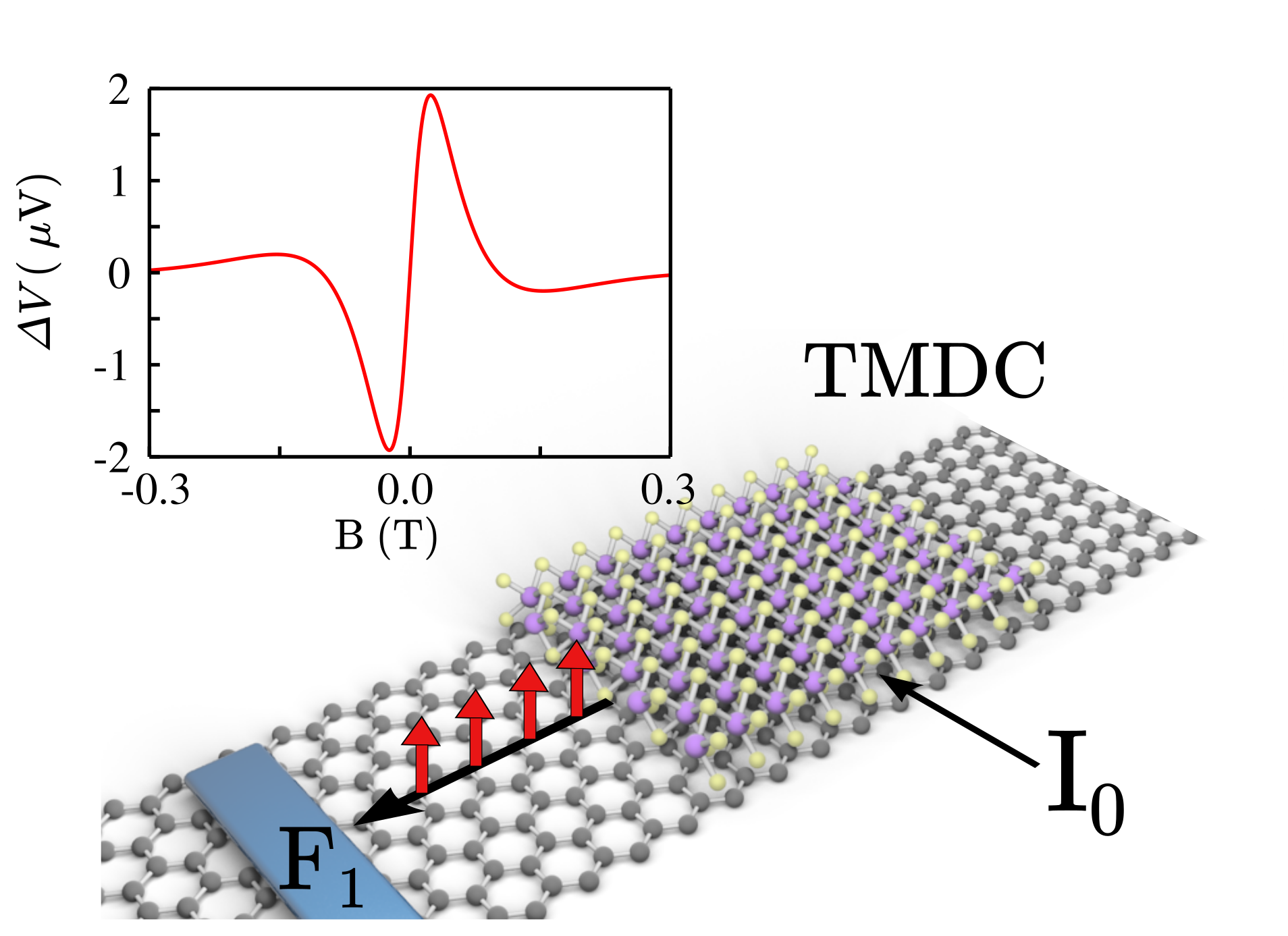}
\caption{Proposed setup for measuring the intrinsic SHE. A current is passed through the graphene/TMDC heterostructure, generating an out-of-plane spin current via the spin Hall effect. The inset shows the expected modulation of the nonlocal signal by an in-plane magnetic field. }
\label{fig:spinHalleffectmeas}
\end{figure}

Given that SOC in pristine graphene is very small, one way to avoid such a situation is to place the TMDC only at the detection or injection region, as represented schematically in Fig.\ \ref{fig:spinHalleffectmeas} and initially explored in Ref. \citenum{Yan2017, SaveroTorres2017} for platinum. Because the spin relaxes isotropically in pristine graphene, modulation due to a magnetic field can be described by the standard Hanle theory in Eq.\ (\ref{eq:final-hanlez}) with the substitutions $P^2 \rightarrow \gamma_{\text{sH}} P$ and $\text{Re}\left\{...\right\} \rightarrow \text{Im}\left\{...\right\}$. The inset of Fig.\ \ref{fig:spinHalleffectmeas} shows the expected Hanle modulation with a magnetic field applied in the propagation direction. Using typical experimental parameters \cite{Benitez2017}, a modulation of 2 $\mu$V is obtained, which is within the experimentally measurable range.

%----------------------------------------FIGURE-----------------------------------------%
%------------------------------------------------------------------------------------------%

%%%%%%%%%%%%%%%%SUBSECTION%%%%%%%%%%%%%%%%%
\subsection{Rashba-Edelstein effect}
%%%%%%%%%%%%%%%%%%%%%%%%%%%%%%%%%%%%%%%
So far we have discussed the possibility of using the SHE for the electrical generation of spins. However, there is another spin-orbit-based phenomenon that could be also used for this purpose, the current-induced spin polarization (CISP). This phenomenon, also called the inverse spin galvanic effect or Rashba-Edelstein effect, consists of generating a nonequilibrium spin density from the imposition of an electrical current \cite{Inoue2003, Manchon2009, MihaiMiron2010, Gambardella2011, Garello2013}. A simplified picture of the CISP can be understood as follows: when an external electric field accelerates the electrons in a material with spin-orbit coupling, the effective spin-dependent magnetic field from the SOC will rotate due to the change in the momentum of the electron. At steady state, the electrons' spins will be aligned with the new direction, thus producing a macroscopic spin polarization. 

However, not every type of spin-orbit coupling will produce such an effect. In general, what is required is an in-plane momentum-dependent spin texture. This is why the effect is expected in Rashba systems, where spin-momentum locking forces the spin-texture to be in-plane, and the nonequilibrium spin polarization is formed perpendicular to the direction of the electrical current. For such systems, the nonequilibrium spin density $n_s^\parallel$ can be generically described by
\begin{equation}
{\delta n_s^\parallel} = -\frac{e E_x \varepsilon_F}{v_\text{F} \hbar^2} \tau_p \mathcal{F}(\varepsilon_F,\lambda_\text{R},\lambda_{\text{VZ}}),
\end{equation}
where $\varepsilon_\text{F}$ is the Fermi energy and $\mathcal{F}(\varepsilon_F,\lambda_\text{R},\lambda_{\text{VZ}})$ is a dimensionless function whose form depends on the scattering mechanism and the type of spin-orbit coupling \cite{Dyrdal2014, Offidani2017}.

\begin{figure}[h!]
\centering
\includegraphics{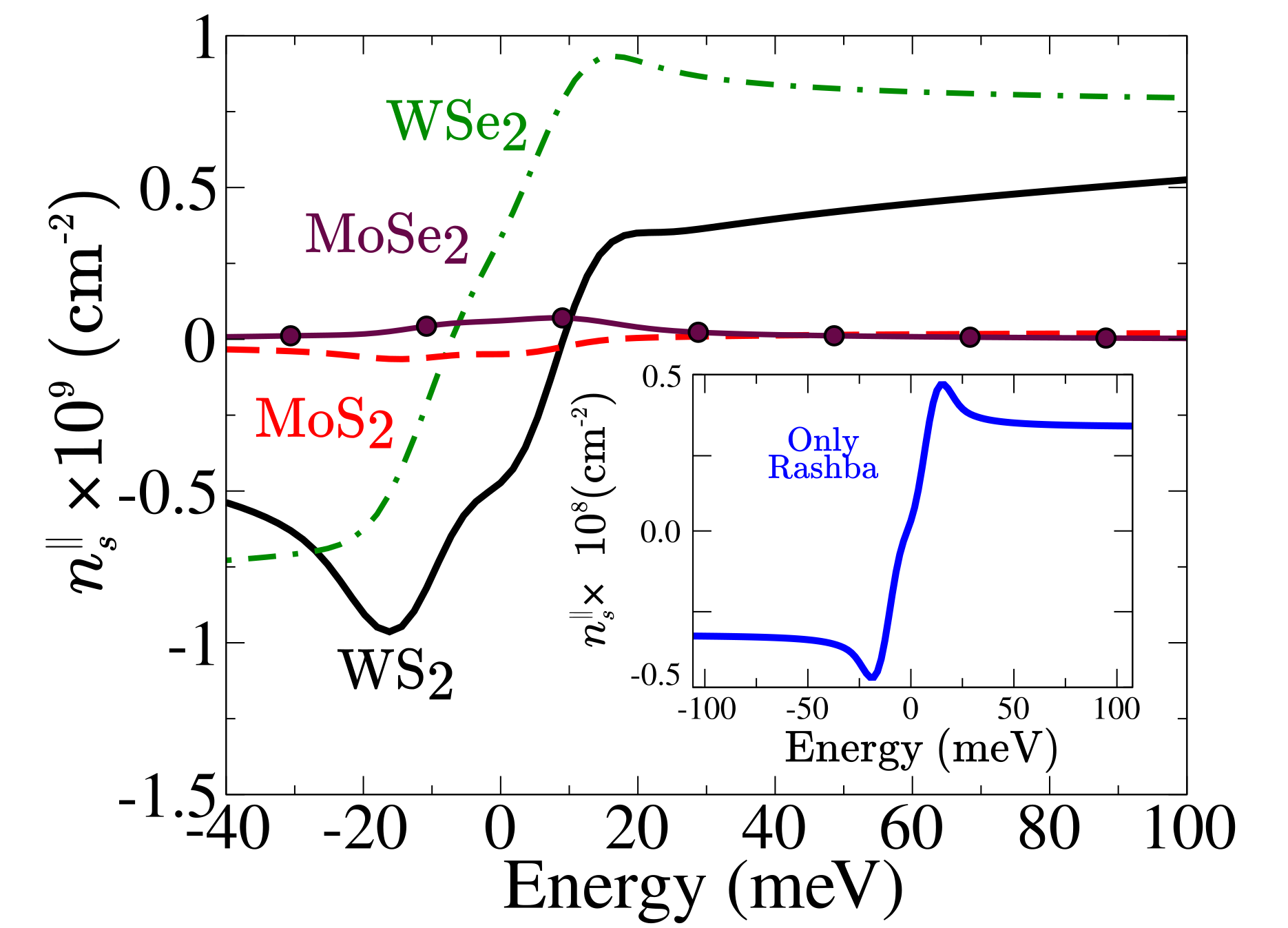}
\caption{Nonequilibrium spin density induced by an electric field of 1V/$\upmu$m for different graphene/TMDC heterostructures, computed with the Kubo-Bastin formula. In the inset we show the case with only the Rashba parameter of WS$_2$. Graphene/TMDC model parameters are shown in Table \ref{table_params}.} \label{spinacc}
\end{figure}

The formation of a steady state nonequilibrium spin density relies on the scattering-induced damping of the spin precession and the establishment of the diffusive regime, and thus care should be taken when considering the clean limit. Nonetheless, the existence of a finite nonequilibrium spin density in the clean limit is a prerequisite for its formation in the diffusive regime, because it is an indication that the spin texture induced by the spin-orbit coupling is capable of polarizing the spins. In Fig.\ \ref{spinacc} we show the nonequilibrium spin density for graphene/TMDC heterostructures, computed using Eq.\ (\ref{Kubo-Basin}) with the replacement $j_{\beta}^z \rightarrow s_y$. We also show the calculation of the pure Rashba system using the graphene/WS$_2$ Rashba parameter, with the simulation nicely reproducing the analytical prediction in this regime \cite{Dyrdal2014}. Despite having the same Rashba field, the nonequilibrium spin density for the full graphene/WS$_2$ model is one order of magnitude larger, which comes from the combination of Rashba and valley-Zeeman SOC. Similar values are obtained for the other heterostructures, which places them, in terms of the CISP effect, next to other strong-SOC metals. Such preliminary results agree quite well with the prediction of Offidani et al.; an optimal spin-to-charge conversion in graphene/TMDCs even in the presence of disorder \cite{Offidani2017}.

In presence of disorder, what matters is how much nonequilibrium spin density is generated
for a given charge current density $j_x$. The figure of merit proposed to quantify the charge-to-spin conversion efficiency is given by \cite{Offidani2017}
\begin{equation}
\varrho= \frac{2 v_F \delta n_s^\parallel}{j_x} .
\end{equation}

This quantity takes values between zero and one, with the maximum representing full efficiency. It was recently shown that for Dirac-Rashba systems, this quantity is close to unity inside the Rashba pseudogap\footnote{The Rashba pseudogap is the range of energies where only a single band, with well defined spin helicity, crosses the Fermi level. For Dirac-Rashba systems is located between zero and $2\lambda_\text{R}$ } and decreases as a power law for higher energies \cite{Offidani2017}. In the same work, it was also demonstrated that as long as the Rashba spin-orbit coupling is the dominant interaction, the large charge-to-spin conversion remains robust to disorder, spin-valley coupling, and temperature. When valley-Zeeman SOC begins to play a more dominant role, the tilting of the spin texture will effectively reduce the in-plane component, and can thus be detrimental for the charge-to-spin conversion. Although \emph{a priori} the theory of charge-to-spin conversion should remain valid, this is a situation that remains to be evaluated.

%----------------------------------------FIGURE-----------------------------------------%
%------------------------------------------------------------------------------------------%

%%%%%%%%%%%%%%%%SECTION%%%%%%%%%%%%%%%%%
\section{Conclusions} \label{sec_conclusions}
%%%%%%%%%%%%%%%%%%%%%%%%%%%%%%%%%%%%%
In this review, we presented an introduction to the theory of spin-orbit effects induced in graphene by TMDCs, and their connection to measurements of Hanle precession, weak antilocalization, and the spin Hall effect. The common thread running through all of these topics is the presence of valley-Zeeman spin-orbit coupling, which significantly alters the behavior of spin transport in these systems by tying the in-plane spin dynamics to intervalley scattering processes.  This leads to a giant anisotropy in the spin relaxation, with in-plane spins relaxing much faster than out-of-plane spins. As shown in Section \ref{spin-precession}, this anisotropy can be determined experimentally with variations on the traditional Hanle measurement. The valley-Zeeman SOC also plays an important role in weak antilocalization, as it appears to be responsible for the very short symmetric spin-orbit times that are extracted from magnetoconductivity measurements. Finally, our calculations have indicated that valley-Zeeman SOC can significantly enhance the intrinsic spin Hall effect in graphene. However, this effect is destroyed by intervalley scattering, meaning that it should only be measurable in systems without a sharp WAL signature and no spin lifetime anisotropy.

\begin{figure}[h!]
\centering
\includegraphics[width=1\columnwidth]{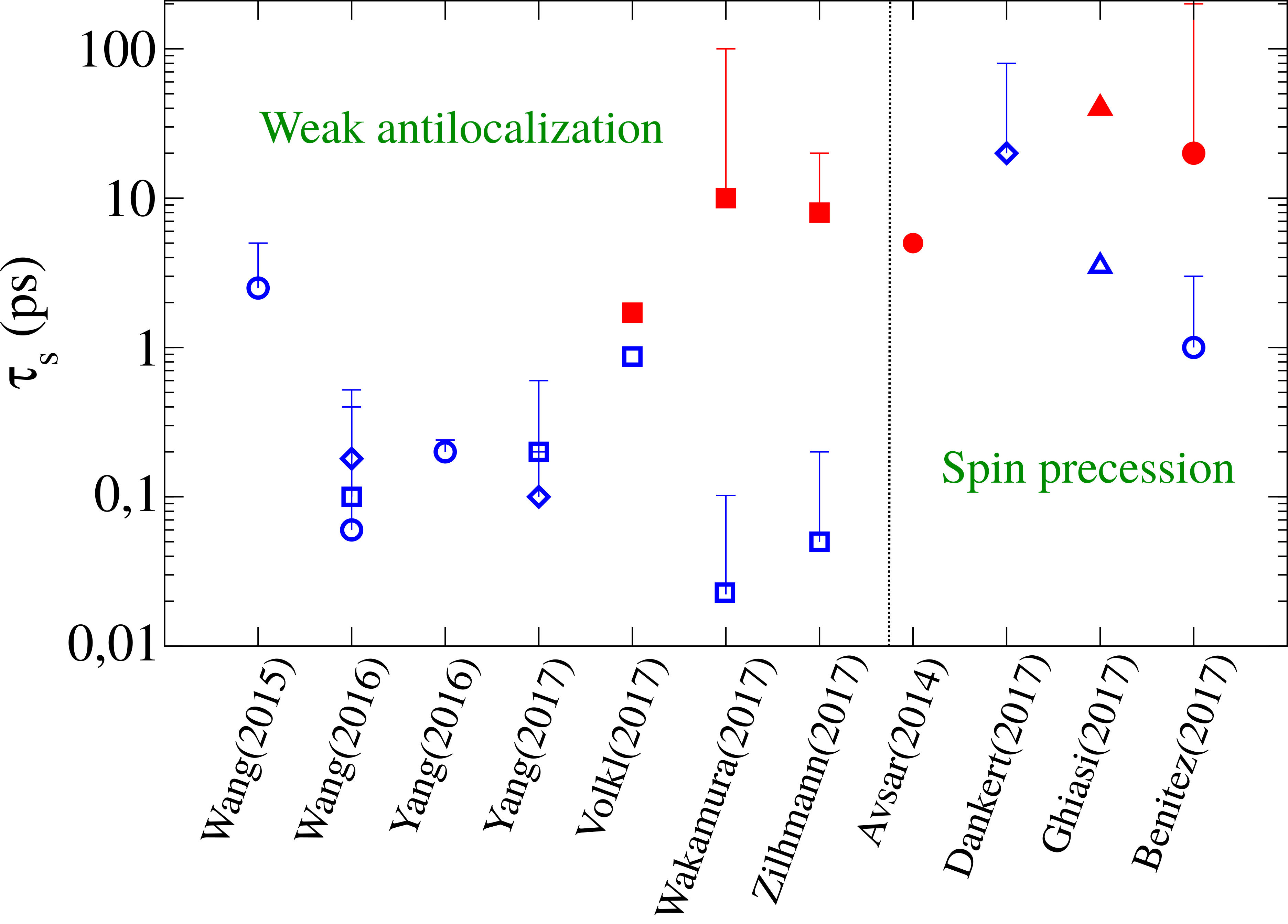}
\caption{Summary of the in-plane (open blue symbols) and out-of-plane (filled red symbols) spin relaxation times for different graphene/TMDC heterostructures, extracted from WAL and Hanle measurements. Squares, circles, diamonds, and triangles correspond to graphene interfaced with WS$_2$, WSe$_2$, MoS$_2$, and MoSe$_2$, respectively. The error bars denotes the variation over different momentum relaxation times.}
\label{fig:spinlifetimesummary}
\end{figure}

In Fig.\ \ref{fig:spinlifetimesummary} we present a summary of all the experimental measurements of spin lifetime to date. The measurements reveal three main trends. First, the symmetric spin lifetimes derived from weak antilocalization and the Hanle in-plane spin lifetimes usually differ by an order of magnitude, which is something our theory currently does not capture, and which calls for further theoretical development as well as new measurements of both quantities within the same sample. Second, the spin lifetime anisotropy can vary from tens to hundreds in different experiments; this is not surprising given that different devices exhibit different disorder and interface quality, which would impact the intervalley scattering rates and SOC strength, respectively. Third, the spin lifetime of graphene is strongly suppressed which, in combination with the giant spin lifetime anisotropy, is fundamental evidence of proximity-induced spin-orbit coupling. We emphasize that interface quality and cleanliness of device fabrication, which can be enhanced by using hBN to encapsulate the heterostructures \cite{Mayorov2011, Geim2013}, can play a crucial role. Additionally, more characterization would help to understand the discrepancies between Hanle and WAL results. In fact, not many experiments carry out an extensive interface characterization; Raman spectroscopy \cite{Li2017}, photoluminescence mapping \cite{Yang2017}, and scanning probe microscopy \cite{Quang2017} would complement the in-plane electrical measurements typically carried out.

%----------------------------------------FIGURE-----------------------------------------%
%---------------------------------------------------------------------------------------%

It is important to note that all our analyses have been performed considering scalar impurities and uniform SOC fields. However, there are experimental results suggesting that the SHE in graphene/TMDC heterostructures may also originate from chalcogenide vacancies \cite{Avsar2014}. These vacancies may constitute a kind of short-range disorder leading to strong intervalley scattering, thus suppressing the intrinsic SHE. However, they also possess a strong local SOC field which could then produce the extrinsic spin Hall effect. Therefore, by carefully tuning the concentration of vacancies, one could observe a transition from intrinsic to extrinsic spin Hall effect. If their local SOC is of the valley-Zeeman type, they could also be responsible for the anisotropy seen in the Hanle experiments, as well as the small symmetric spin-orbit times extracted from measurements of weak antilocalization. In the reduced WAL analysis of Eq. (\ref{eq_wal}), the spin relaxation rates arising from uniform and nonuniform SOC are lumped together in both $\tau_\text{sym}^{-1}$ and $\tau_\text{asy}^{-1}$, but considering the full WAL formula may allow one to separate these terms.

Finally, very recent theoretical studies of bilayer graphene (BG) deposited on TMDC have appeared. Because of the different band structure of BG with respect to graphene \cite{McCann2013}, only the SOC of the valence band is increased while leaving the SOC of the conduction band almost unaffected \cite{Gmitra2017}. More importantly, one can swap the SOC strength between these bands with a perpendicular electric field, which could be used to create a spin transistor or enable valleytronic phenomena \cite{Khoo2017}. In the same way, we would like to emphasize that such proximity-induced spin-orbit effects are not limited to TMDCs. In reality, these features are a consequence of broken symmetries in graphene \cite{Kochan2017}, and any high-SOC substrate with broken sublattice symmetry should have a similar impact. In fact, it has recently been predicted that graphene on a topological insulator (TI) possesses very similar features \cite{Kenan2017}, opening the door to a possibly rich interplay between graphene with proximity-induced SOC and TI surface states.

%%%%%%%%%%%%%%%%SECTION%%%%%%%%%%%%%%%%%
\section*{Acknowledgement}
%%%%%%%%%%%%%%%%%%%%%%%%%%%%%%%%%%%%%

ICN2 is supported by the Severo Ochoa program from Spanish MINECO (Grant No. SEV-2013-0295) and funded by the CERCA Programme / Generalitat de Catalunya. The authors acknowledge the Spanish Ministry of Economy and Competitiveness and the European Regional Development Fund (Project  No. FIS2015-67767-P MINECO/FEDER), the Secretar\'ia de Universidades e Investigaci\'on del Departamento de Econom\'ia y Conocimiento de la Generalidad de Catalunya (2014 SGR 58), PRACE and the Barcelona Supercomputing Center (Project No. 2015133194), and the European Union Seventh Framework Programme under Grant Agreement No. 696656 Graphene Flagship.We thank Williams Savero, Antonio Benitez, Fr\'ed\'eric Bonell, Juan F. Sierra, Marius V. Costache, Aires Ferreira, Branislav K. Nikolic, Simon Zihlmann, P\'eter Makk  and Sergio Valenzuela for the useful discussions. H\'el\`ene Bouchiat and Taro Wakamura are also thank for kindly providing their experimental data

%%%%%%%%%%%%%%%%SECTION%%%%%%%%%%%%%%%%%
\section*{Conflict of interest}
%%%%%%%%%%%%%%%%%%%%%%%%%%%%%%%%%%%%%
There are no conflicts to declare.

\end{document}